\documentclass[
	aps, pra, twocolumn,superscriptaddress,
	floatfix, 
    nofootinbib,longbibliography,
	tightenlines,
	10pt
]{revtex4-2}
\usepackage[final]{graphicx}
\usepackage{times,bbm,amsmath,amssymb}
\usepackage{epsfig,color}
\usepackage{xcolor}
\usepackage{hyperref}
\hypersetup{
    colorlinks = true
}
\usepackage{cleveref}

\usepackage{float,siunitx}
\usepackage[caption = false]{subfig}

\usepackage[greek,english]{babel}
\usepackage{thumbpdf,enumerate}
\usepackage{booktabs}
\usepackage{sidecap}
\usepackage[scaled=.8]{couriers}    
\usepackage{pstricks}
\usepackage{multirow}
\usepackage{placeins}
\usepackage{relsize}  
\usepackage{pst-grad,bm}
\usepackage{epigraph}
\usepackage{gensymb}
\usepackage{longtable}
\usepackage{booktabs}
\usepackage{gensymb}

\usepackage{soul}
\usepackage{ulem} 
\usepackage{acronym}

\usepackage{physics}

\usepackage{tikz}

\usepackage{bbold}
\usepackage{amsthm}
\usepackage{microtype}

\newcommand{\comment}[1]{}

\begin{document}
\title{Certification of Gaussian Boson Sampling via graph theory}

\author{Taira Giordani}
\thanks{These two authors contributed equally}
\affiliation{Dipartimento di Fisica, Sapienza Universit\`{a} di Roma,
Piazzale Aldo Moro 5, I-00185 Roma, Italy}

\author{Valerio Mannucci} 
\thanks{These two authors contributed equally}  
\affiliation{Dipartimento di Fisica, Sapienza Universit\`{a} di Roma,
Piazzale Aldo Moro 5, I-00185 Roma, Italy}

\author{Nicol\`o Spagnolo}
\affiliation{Dipartimento di Fisica, Sapienza Universit\`{a} di Roma,
Piazzale Aldo Moro 5, I-00185 Roma, Italy}

\author{Marco Fumero} 
\affiliation{Dipartimento di Informatica, Sapienza Universit\`{a} di Roma,
Via Salaria 113, I-00198 Roma, Italy}

\author{Arianna Rampini} 
\affiliation{Dipartimento di Informatica, Sapienza Universit\`{a} di Roma,
Via Salaria 113, I-00198 Roma, Italy}

\author{Emanuele Rodolà}
\email[Corresponding author: ]{rodola@di.uniroma1.it}
\affiliation{Dipartimento di Informatica, Sapienza Universit\`{a} di Roma,
Via Salaria 113, I-00198 Roma, Italy}

\author{Fabio Sciarrino}
\email[Corresponding author: ]{fabio.sciarrino@uniroma1.it}
\affiliation{Dipartimento di Fisica, Sapienza Universit\`{a} di Roma,
Piazzale Aldo Moro 5, I-00185 Roma, Italy}

\begin{abstract}
  Gaussian Boson Sampling is a non-universal model for quantum computing inspired by the original formulation of the Boson Sampling problem. Nowadays, it represents a paradigmatic quantum platform to reach the quantum advantage regime in a specific computational model. Indeed, thanks to the implementation in photonics-based processors, the latest Gaussian Boson Sampling experiments have reached a level of complexity where the quantum apparatus has solved the task faster than currently up-to-date classical strategies. In addition, recent studies have identified possible applications beyond the inherent sampling task. In particular, a direct connection between photon counting of a genuine Gaussian Boson Sampling device and the number of perfect matchings in a graph has been established. In this work, we propose to exploit such a connection to benchmark Gaussian Boson Sampling experiments. We interpret the properties of the feature vectors of the graph encoded in the device as a signature of correct sampling from the true input state. Within this framework, two approaches that exploit the distributions of graph feature vectors and graph kernels are presented. Our results provide a novel approach to the actual need for tailored algorithms to benchmark large-scale Gaussian Boson Samplers.  
  \end{abstract}

\maketitle

\section*{Introduction}
Quantum processors and quantum algorithms promise substantial advantages in computational tasks \cite{Harrow2017_supremacy}. Recent experiments have shown significant improvements towards the realization of large scale quantum devices operating in the regime in which classical computers cannot reproduce the output of the calculation \cite{Harrow2017_supremacy,Arute2019, Wu_2021_supremacy, Zhong_GBS_supremacy, zhong2021phaseprogrammable}. 

An intriguing computational problem concerning quantum photonic processors is Boson Sampling (BS) and, more recently, its variant Gaussian Boson Sampling (GBS). The BS paradigm corresponds to sampling from the output distribution of a Fock state with $n$ indistinguishable photons after the evolution through a linear optical $m$-port interferometer \cite{AA, Brod19review}. This problem turns out to be intractable for a classical computer, while a dedicated quantum device can tackle such a task towards unequivocal demonstration of quantum computational advantage. The GBS variant replaces the quantum resource of the BS, i.e the Fock state, with single-mode squeezed vacuum states (SMSV). This change to the original problem enhances the samples generation rate with respect to BS performed with probabilistic sources, and preserves the hardness of sampling from a quantum state \cite{Lund_SBS,wcqoscct,Hamilton2017, Deshpande_GBS_th_supremacy}. 

The GBS problem has drawn attention for the practical chance to achieve the quantum advantage regime. After the small scale experiments \cite{Zhong19, Paesani2019, thekkadath2022experimental}, 
the latest GBS instances have just reached the condition where the quantum device has solved the task faster than current state-of-the-art classical strategies \cite{Zhong_GBS_supremacy, zhong2021phaseprogrammable}. The interest in GBS also concerns applications for sampling for gaussian states beyond the original computational advantage. The probability of counting $n$-photon in the output ports of a GBS is proportional to the squared \emph{hafnians} of an appropriately constructed matrix, that takes into account the unitary transformation $U$ representing the optical circuit and the covariance matrix of the input state. Computing Hafnians of a matrix is as hard as computing \emph{permanents} that describe the amplitude of the BS output states. The hafnians have a precise interpretation in graph theory since their calculation corresponds to counting the perfect matchings in a graph. The adjacency matrix of a graph can be encoded in a GBS, and then the collected samples are informative about the graph properties. Recently, GBS-based algorithms for solving well-known problems in graph theory have been formulated \cite{Arrazzola_densesubgraph, Shuld_GBS_graphsimilarity, Bradler_2021} and tested in a first proof-of-principle experiment of the GBS within a reconfigurable photonic nano-chip \cite{Arrazola2021}.

\begin{figure}[t]
    \centering
    \includegraphics[width=\columnwidth]{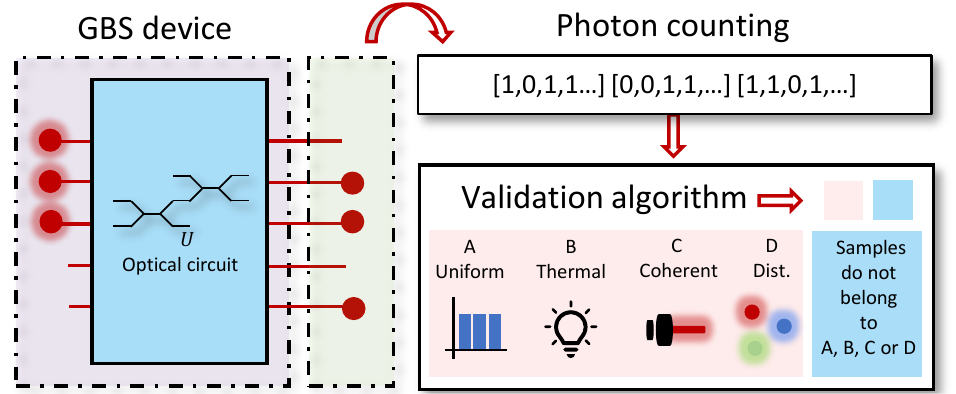}
   \caption{\textbf{Gaussian Boson Sampling validation}. In the GBS paradigm $n$-photon configurations are sampled at the outputs of an optical circuit with $m$ ports. The aim of a validation algorithm is to exclude that the obtained samples could have been generated by classically-simulable models. Previous experiments mainly focused in techniques capable to rule out the uniform sampler, the thermal, coherent and distinguishable SMSV states hypotheses.}
\label{fig:validation}
\end{figure}

These results on the BS framework are thus bringing back photonic platforms as a promising approach to implement quantum algorithms. In parallel, this development is currently accompanied by research efforts aimed at identifying suitable and efficient strategies for system certification. This is indeed a crucial requirement, both for benchmarking quantum devices reaching the quantum advantage regime, as well as validating the operation of such systems whenever they are employed to solve specific computations. While several methodologies have been developed and reported, certification of quantum processors is still an open problem. In the case of BS and GBS, direct calculation or sampling from the output distribution cannot be performed efficiently by classical means, and are thus not viable for large-scale implementations with many photons and ports of the optical circuit \cite{clifford2017classical,Neville2017,Quesada_exact_simulation,Quesada_exact_simulation_speedup, Bulmer_Markov_GBS}. Then, it is preferable to switch the problem towards a validation approach, i.e to exclude that the samples could be reproduced by specifically chosen classical models. The validation tests first developed for the BS problem focus on ruling out the uniform sampler, the distinguishable particle sampler and the mean-field sampler hypotheses \cite{Aaronson14, Tichy, Spagnolo2, Carolan15, Crespi16,Viggianiello18, Walschaers16, Giordani18, agresti2019pattern, FlaminiTSNE}. Recent efforts have been also dedicated to addressing partial photon distinguishability \cite{Viggianiello17optimal, Renema_partial_2020}, which is a crucial requirement that can spoil the complexity of the computation \cite{Renema_2018_classical, Moylett_2019}. This validation approach, originally conceived for the BS problem and based on defining suitable alternative hypotheses, has been subsequently extended to the GBS variant (see Fig.\ref{fig:validation}). In various experiments, the samples from GBSs have been validated against alternative classically-simulable hypotheses, such as the thermal, coherent and distinguishable SMSV states \cite{Zhong19, Paesani2019, Zhong_GBS_supremacy, zhong2021phaseprogrammable}. 
These GBS validation examples include variations of Bayesian approaches \cite{Zhong_GBS_supremacy, zhong2021phaseprogrammable} or algorithms based on the statistical properties of two-point correlation functions \cite{Walschaers16, Giordani18} that can be used also for GBS to exclude thermal and distinguishable SMSV samplers \cite{hanbury}. In~\cite{zhong2021phaseprogrammable} a more refined analysis investigates the possibility of describing the experimental results as lower order interference processes, thus not involving all the generated photons. This approach is strictly related to sampling algorithms based on low-order interference approximations \cite{popova2021cracking} or low-order marginal probabilities
\cite{AA, Renema_partial_2020, renema2020marginal, villalonga2021efficient}.
In parallel, studies regarding the classical simulability of BS and GBS in terms of photon losses have also been carried out \cite{Oszmaniec_2018,GarciaPatron2019simulatingboson,Brod2020classicalsimulation, Qi_lossyGBS}.

Besides these examples of GBS validations, there is a lack of tailored algorithms for GBS that could be efficient in the regime of quantum advantage. In this work, we propose a validation protocol based on the deep connection between GBS and graph theory. We consider the features of the graph extracted from the GBS samples as a signature of the correct sampling from indistinguishable SMSV states. Within this framework, we present two approaches. The first method considers the space spanned by the feature vectors extracted from photon counting samples obtained from different gaussian states. Then, a classifier, such as a neural network, can be trained to identify an optimal hyper-surface to distinguish a true GBS and the mock-up hypotheses in this space. The second method investigates the properties of the kernel generated by the feature vectors of each class of gaussian state.  
Both approaches exploit macroscopic quantities that can be retrieved in a reasonable time from the measured GBS samples.

This work is organized as follows. First, we review the concept of sampling from gaussian state of light and the relationship with counting graph perfect matchings. Then, we present the validation methods based on the properties of graph feature vectors and kernels. We conclude by providing insights on the effectiveness of the proposed approach to discriminate genuine GBS from different alternative hypotheses.

\section*{Gaussian Boson Sampling and its connection with graphs}
\begin{figure*}[ht]
    \centering
    \includegraphics[width = \textwidth]{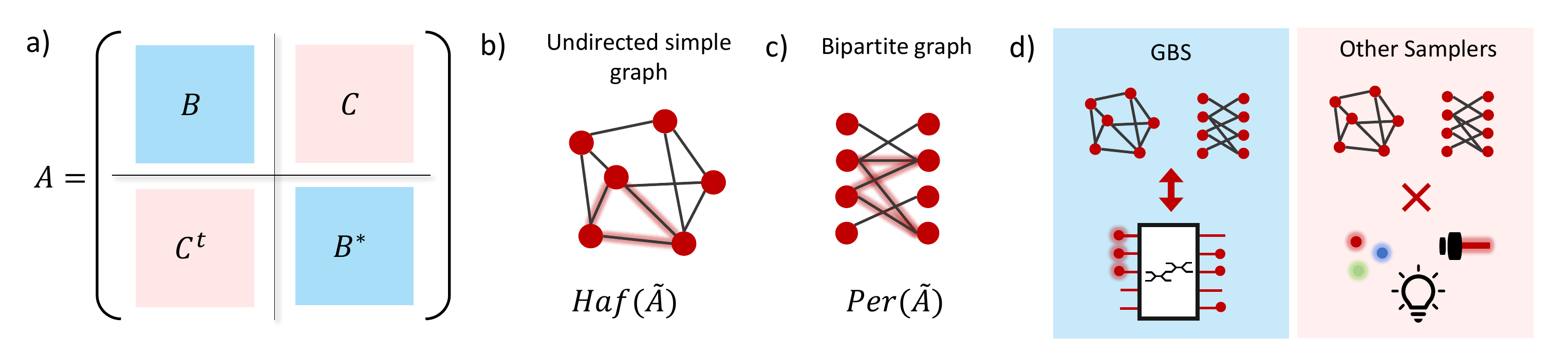}
    \caption{\textbf{Gaussian Boson Sampling and graph perfect matching.} a) Structure of the $2m\times 2m$ sampling matrix for $m$ independent gaussian states injected in a $m$-port interferometer. b) The hafnian as the operation to count the perfect matchings in a simple undirected graph with $n$ nodes. c) The permament as the same operation for a bipartite graph. d) The adjacency matrix of undirected graphs, even in the bipartite case, can be encoded in GBS devices. }
    \label{fig:haf_perm}
\end{figure*}
\vspace{1ex}\noindent\textbf{Background.}
Here we briefly review the general theory of the probability to obtain $n$-photon configurations from a set of indistinguishable gaussian input states $\rho_i$ such that $\rho_{in}=\otimes_{i=1}^m \rho_i$, and  distributed in $m$ optical modes, after the evolution in a multi-port interferometer. 
Given the $2m\times2m$ covariance matrix $\sigma$ that identifies the gaussian state, and the output configuration $\vec{n}= (n_1, n_2, \dots, n_m)$, where $n_i$ is the number of photons detected in the output port $i$ such that $\sum_{i=1}^m n_i = n$, we have  
\begin{equation}
\text{Pr}(\vec{n})=|\sigma_Q|^{-\frac{1}{2}}\frac{\text{Haf}\,(A_{\vec{n}})}{\prod_{\vec{n}}{n_i}!} \,.
\label{eq:hafn_gbs}
\end{equation}
The quantity $\sigma_Q$ is $\sigma+\frac{1}{2}\mathbb{I}_{2m}$ where $\mathbb{I}_{2m}$ is the $2m\times2m$ identity operator; $A_{\vec{n}}$ is a sub-matrix of the overall matrix $A$ that contains the information about the optical circuit represented by the transformation $U$ and the covariance of the input state, while $\text{Haf}$ stands for the hafnian of the matrix. More precisely, $A_{\vec{n}}$ is the $n \times n$ sub-matrix obtained by taking $n_i$ times the $i$-th row and the $i$-th column of $A$ \cite{Hamilton2017, DetailedstudyGBS}. The hafnian of $A_{\vec{n}}$ corresponds to the summation over the possible perfect matching permutations, i.e, the ways to partition the index set $\{1,\cdots ,n\}$ into $n$/2 pairs such that each index appears only in one pair (see also~\cite{Caianiello1953}).
The hafnian is in the \#P-complete complexity class, and is a generalization of the permanent of a matrix $M$ according to the following expression:
\begin{equation}
    \text{Per}(M)=\text{Haf}\begin{pmatrix}
    0 & M\\
    M^t & 0
    \end{pmatrix}\,.
    \label{eq:permanent}
\end{equation}
The above description has been used to define a classically-hard sampling algorithm, using indistinguishable SMSV states with photon-counting measurements
\cite{wcqoscct, Hamilton2017, DetailedstudyGBS}. More specifically, in Fig. \ref{fig:haf_perm}a we report the structure of the sampling matrix $A$ for an input state $\rho$ that has zero displacement. In the language of quantum optics the displacement is the operation that generates a coherent state from the vacuum. Then, the blocks $B$ and $C$ highlighted in the Fig. \ref{fig:haf_perm}a correspond to the contribution of squeezed and thermal light respectively in the input state. Pure, indistinguishable, SMSV states display a $C=0$ and $B=U \text{diag} (\tanh{s_1}, \dots, \tanh{s_m}) U^t$, where $s_i$ are the squeezing parameters of each $\rho_i$ \cite{DetailedstudyGBS}. According to this representation the expression in Eq. \eqref{eq:hafn_gbs} becomes
\begin{equation}
    \text{Pr}(\vec{n})_{\text{SMSV}} = |\sigma_Q|^{-\frac{1}{2}}\frac{|\text{Haf}\,({B}_{\vec{n}})|^2}{\prod_{\vec{n}}{n_i}!}\,,
    \label{eq:smsv_states}
\end{equation}
where ${B}_{\vec{n}}$ is the submatrix of $B$ obtained from the string $\vec{n}$ as described at the beginning of the section.  

\vspace{1ex}\noindent\textbf{Connection to graph theory.}
Recently, several works have identified a connection  
between the GBS apparatus and graph theory \cite{ArrazolaQOpt,Arrazzola_densesubgraph, Bradler_2021}. These studies take advantage of such a relationship to formulate GBS-based algorithms in the context of graph-similarity and graph kernels.
The algorithms exploit the fact that the vectors extracted from GBS samples can be considered a feature space for a graph encoded inside the apparatus. In particular, they are strictly correlated to a class of classical graph kernels that count the number of $n$-matchings, i.e., the perfect matchings of the sub-graph with $n$ links in the original graph encoded inside the GBS. Given $A$ the adjacency matrix of the graph, the number of perfect matchings 
is proportional to the hafnian of the matrix, thus corresponding to the output probabilities in Eqs. \eqref{eq:hafn_gbs} and \eqref{eq:smsv_states}.   
 Indeed, any symmetric matrix, such as the graph adjacency matrices, can be decomposed accordingly to the Takagi-Autonne factorization as $A = U \text{diag}(c\lambda_1,\dots, c\lambda_m) U^{t}$, where $\lambda_i$ are real parameters in the range $[0,1]$, $c$ is a scaling factor and $U$ is a unitary matrix. This decomposition matches with the expression of the sampling matrix $B$ of SMSV states when $\lambda_i = \tanh{s_i}$. Also squeezed states with very small displacement have a $n$-photon probability distribution that can be expressed through hafnians.  
 For example, displaced squeezed states have been investigated in the context of graph similarity, where a small amount of displacement has been employed as a hyper-parameter to enhance the graphs' classification accuracy \cite{Shuld_GBS_graphsimilarity}. 
 
Regarding the sub-matrix selected by the sampling process, the configuration $\vec{n}$ identifies the elements of the sub-matrix $A_{\vec{n}}$ that represent an induced sub-graph (see Fig \ref{fig:haf_perm}b). The nodes of the original graph $A$ corresponding to detectors with zero counting are deleted, together with any edges connecting these nodes to the others.
If some elements $n_i$ of $\vec{n}$ are larger than one, i.e. these detectors count more than one photon, $A_{\vec{n}}$ describes what we call an \emph{extended induced sub-graph} in which the corresponding nodes and all their connections are duplicated $n_i$ times.

It is worth noting that also the permanent has a precise meaning in the context of graphs. Indeed, the matrix on the right-hand side of Eq. \eqref{eq:permanent} corresponds to the adjacency matrix of a bipartite graph. In other words, the permanent calculation provides the number of perfect matchings for this class of graphs (see Fig. \ref{fig:haf_perm}c). One may ask whether other sampling processes regulated by permanent calculations, such as the BS and the thermal samplers (see Appendix \ref{app:sampling}), could have a relationship with bipartite graphs. 
The BS output distribution is defined by the permanent of the sub-matrix from the unitary transformation $U$ representing the circuit. It is clear that not all graphs can be represented by a unitary adjacency matrix. Furthermore, in the BS paradigm, the sub-matrix selected by the sampling process depends also on the input state. This implies that the resulting sub-graph could not have the same symmetries and properties as the original encoded in the $U$ matrix. The latter issue can be overcome by using thermal light, where only the output configuration $\vec{n}$ determines the sub-matrix. However, also for thermal light, the sampling matrix $C$ does not in general represent an adjacency matrix, thus preventing the possibility of encoding any bipartite graphs. 
In conclusion, the GBS devices with squeezed states are the only ones that have a direct connection with graphs (see Fig. \ref{fig:haf_perm}d). 

\section*{Feature vector-based validation algorithm}
In the following, we illustrate two validation algorithms tailored for GBS. The idea behind our protocols is to exploit the connection between the samples of a genuine GBS and the graph properties encoded in the device. 

According to Eq. \eqref{eq:smsv_states} the most likely outcomes from the GBS are those with the highest hafnians, i.e. the output configurations that identify the sub-graph $A_{\vec{n}}$ with the largest number of perfect matchings.
However, we remind that the calculation of a single hafnian is a \#P-complete problem as the counting of the perfect matchings in a graph. Furthermore, estimation of the output probabilities from the quantum devices becomes unfeasible for large system sizes, and thus any protocol should not rely on this ingredient. Then, it is necessary for a successful validation algorithm to exploit quantities that do not depend 
on the evaluation of the probability of a single $\vec{n}$, which would require exponential time for its estimation.

\vspace{1ex}\noindent\textbf{Feature vectors.}
%
It is possible to extract properties from a graph summarized in the so called feature vectors. In the GBS-based algorithms the features of the graph are extracted from a coarse-graining of the output configuration states. For instance, the probability to detect configurations $\{\vec{n}\}$ with $n_i = \{0,1\}$ is linked to the number of perfect matchings of the sub-graphs $\{A_{\vec{n}}\}$ of $A$ which do not have repetition of nodes and edges. Accordingly, the probability of the set of $\{\vec{n}\}$ with two photons in the same output will be connected to the perfect matching in sub-graphs with one repetition of a pair of nodes and edges. The collections of output configurations that identify a family of sub-graphs with a certain number of nodes and edges repetitions are called \emph{orbits} \cite{Shuld_GBS_graphsimilarity}. Given $n$ the total number of post-selected photons in the output, the orbit $O_{\vec{n}}$ is defined as the set of the possible index permutations of  $\vec{n}$. In this work we consider the orbit $O_{[1,1,\dots,1,0 \dots  0]}$ that corresponds to output states with one or zero photon per mode; the orbit $O_{[2,1,\dots,1,0 \dots, 0]}$ that is the collection of the outputs with one mode occupied by two photons and $O_{[2,2,1\dots,1,0 \dots, 0]}$ with two distinct outputs hosting two photons. The graph feature vector components are identified by the probability of each orbit, defined as $Pr(O_{\vec{n}})=\sum_{\vec{n} \in O_{\vec{n}}}Pr(\vec{n})$. In the rest of this work we will refer to the probabilities of the orbits $O_{[1,1,\dots,1,0 \dots  0]}$, $O_{[2,1,\dots,1,0 \dots, 0]}$ and $O_{[2,2,1\dots,1,0 \dots, 0]}$ as $[1, \dots, 1]$, $[2,1, \dots1]$ and $[2,2,1,\dots,1]$ respectively.
    
The orbit probabilities can be estimated directly from photon counting measurements. This method can be applied in GBS experiments. In numerical simulation, direct sampling of photon counting is a viable approach for deriving orbit probabilities of gaussian states that can be sampled classically, such as distinguishable SMSV, thermal and coherent states (see Appendix \ref{app:sampling}). These states reproduce the scenarios that could occur in the experimental realizations of GBS devices. For example, photon losses turn the squeezed light into thermal radiation, while mode-mismatch, such as spectral and temporal distinguishability, breaks the symmetry of boson statistics. Exact estimation of the orbits for indistinguishable SMVS states can be performed by directly calculating all the hafnians, thus requiring evaluation of a large number of complex quantities. A different approach can be employed, based on approximating the orbits probability by a Monte Carlo simulation \cite{Killoran2019strawberryfields, Bromley_2020}. The outputs $\vec{n}$ within an orbit are selected uniformly at random and their exact probabilities are calculated. Then, the probability of the whole orbit after $N$ extractions can be approximated by $Pr(O_{\vec{n}}) \approx \frac{|O_{\vec{n}}|}{N}\sum_{i=1}^N Pr(\vec{n}_i)$, where $|O_{\vec{n}}|$ is the number of elements in the orbit. The adopted strategies reproduce the experimental conditions in which the orbits probabilities are estimated on a finite number $N$ of samples. The code for generating GBS data included routines from Strawberry Fields \cite{Killoran2019strawberryfields} and The Walrus \cite{Gupt2019} Python libraries.
\begin{figure}[t]
    \centering
    \includegraphics[width=\columnwidth]{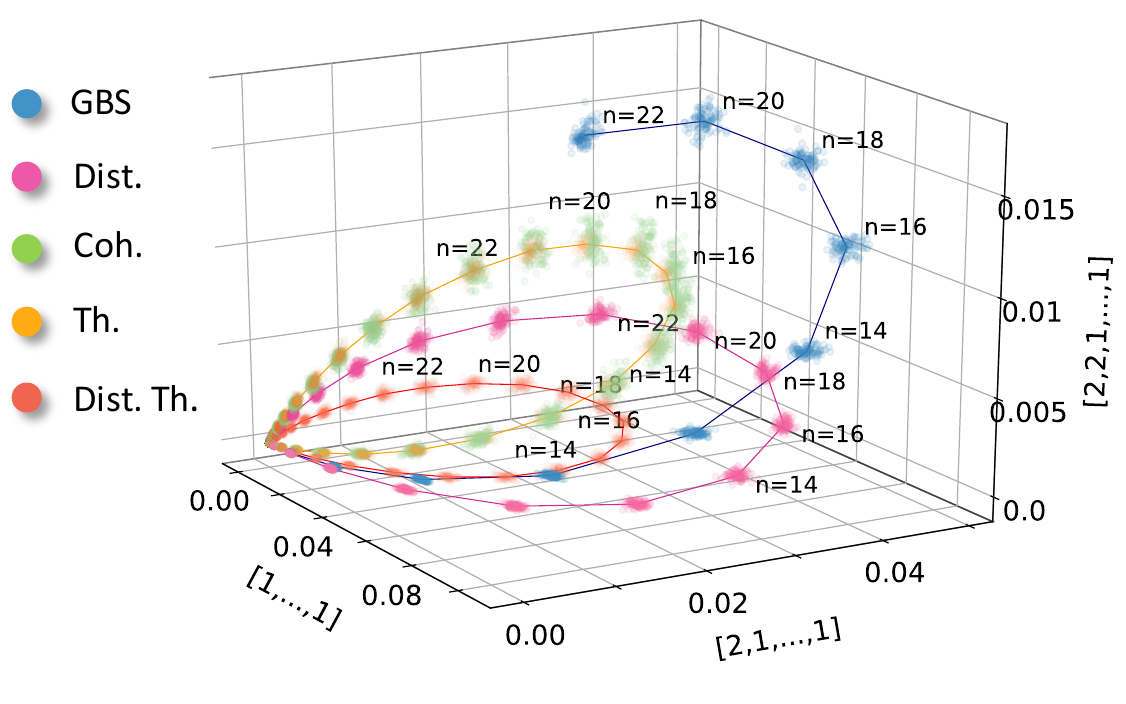}
    \caption{\textbf{Orbits probabilities distribution.} In blue we report the feature vectors for $100$ genuine GBS devices with $m=400$. The squeezers parameters in each GBS were tuned to obtain a photon number distribution centered around $n \sim 16 \ll m$. Each cloud corresponds to the post-selection of different number of photons $n$ in the outputs. This is equivalent to look at the features of $n$-node sub-graphs. In yellow we report the thermal sampler case, in pink the distinguishable sampler, in red the distinguishable thermal sampler and in green the coherent light one. 
    GBS data were generated numerically via Monte Carlo approximation of the orbits probabilities. The maximum size achieved for the simulation corresponds to $n = 22$ for computation time reasons. The data of the other models were extracted from direct sampling of the photon counting. Thermal, coherent and distinguishable thermal samplers display also a non-zero probability to generate odd number of photons.}
    \label{fig:leaves}
\end{figure}

  \begin{figure*}[t] 						    		
\centering
\includegraphics[width = \textwidth]{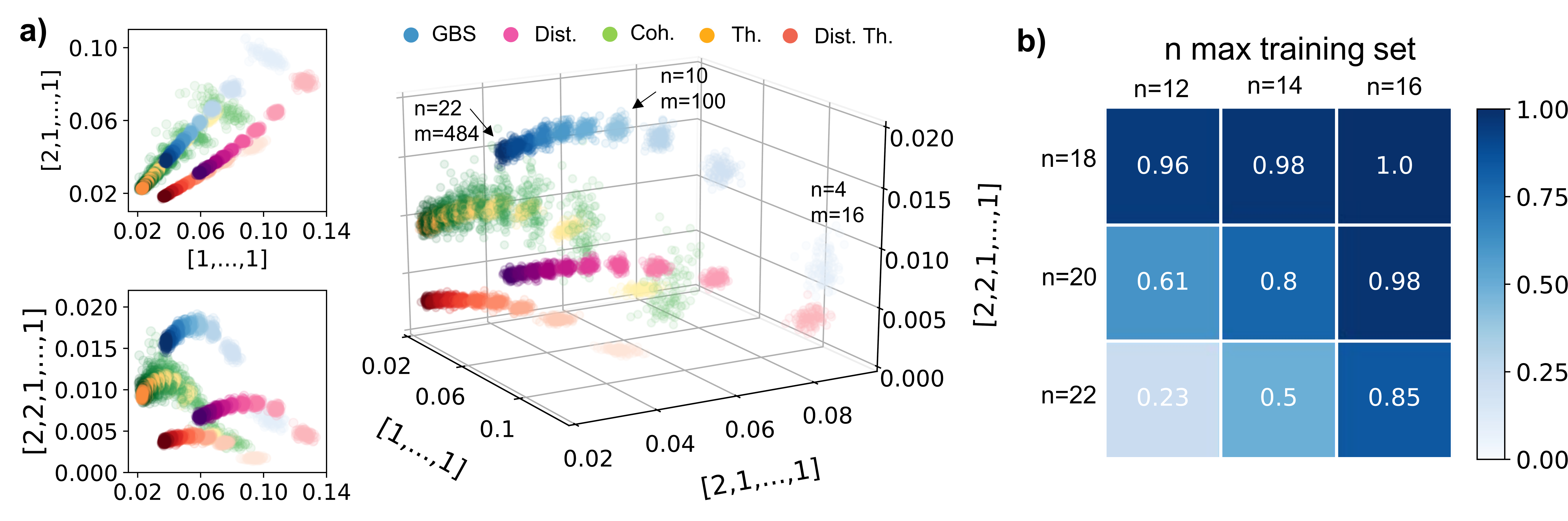}
\caption{\textbf{Orbits probabilities for different sizes of the GBS}. a) Orbits probabilities $[1, \dots, 1]$, $[2, 1, \dots, 1]$, $[2, 2, \dots, 1]$ for different samplers with $n$-photon $\in [4, 6, \dots, 22]$ and $m=n^2$ optical modes. In the blue-scale samples from a genuine GBS device, in yellow data from indistinguishable thermal states, in green the coherent states, in pink the distinguishable SMVS states and in red the distinguishable thermal light. For each $n$ and class of states we sampled $100$ sets of $U$ and $\{s_i\}$. Two orbits are not enough to discriminate the data, while in the space spanned by three orbits the various hypotheses are very well separated. b) Results of the classification accuracy of genuine GBS data by means of a neural network classifier. The network trained with trusted GBS data of smaller sizes indicated along the x axis is able to correctly classify larger GBS devices. }
\label{fig:fv_scaling}
\end{figure*}

\begin{figure*}[t]
    \centering
    \includegraphics[width=\textwidth]{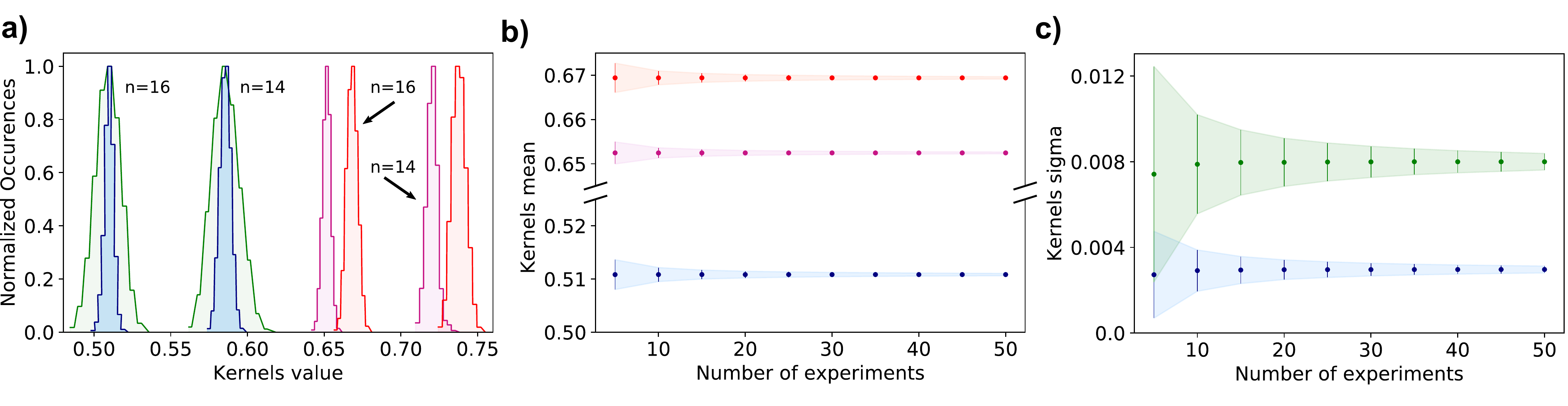}
    \caption{\textbf{Graph kernels distributions.} a) Kernels distributions for the graphs encoded in GBSs with $m$=400 and photon-number distribution centered around $n=16$. The feature vectors have been normalized to a given $n$ and to the space of the three orbits. We report the distributions for the GBS in blue, coherent light sources in green, distinguishable SMSV in pink and distinguishable thermal states in red. The four histograms display different features for any $n$. b-c) Kernels mean and standard deviation for the case $n=16$ and for increasing values of measured graphs. We observe that a small amount of experiments is enough to discriminate genuine GBS kernels. The uncertainties reported in the plots correspond to $3$ standard deviations.}
    \label{fig:kernels}
\end{figure*}

\vspace{1ex}\noindent\textbf{Validation by classification.}
As a first method for validation, we propose the classification of these different samplers in the space spanned by the three feature vector components identified by the orbits $[1, \dots, 1]$, $[2,1,\dots,1]$ and $[2,2,1,\dots,1]$. In Fig.~\ref{fig:leaves} we give an insight of our intuition by reporting an example of the distribution of feature vectors for different graphs and sampler types. The colors underline samples from different models such as genuine GBS, distinguishable SMSV states, coherent light, indistinguishable thermal light emitters and distinguishable thermal states. In this simulation we consider $100$ optical random circuits with $m=400$ modes and $m$ sources set to produce a photon-number distribution centered in $n\ll m$. In this condition we are in the dilute regime where the orbits with low number of photons in the same output have the highest probability to occur. It is worth noting that in this estimation it is necessary to take into account the occurrence of the orbits in the whole space of the GBS, i.e the Hilbert space associated to the all possible $n$-photon states that can be generated by the squeezers. 

Experimentally, such a method requires the knowledge of the photon-number distribution of the sources. Such requirement is not demanding since the characterization of the gaussian sources is a standard preliminary procedure in GBS experiments \cite{Zhong19,Paesani2019, Zhong_GBS_supremacy, Arrazola2021, zhong2021phaseprogrammable}.  Alternatively, the orbits probability can be estimated by post-selecting samples with different total number $n$ and dividing the occurrence of photon counting belonging to the orbit with a given $n$ by the total number of samples. The data of the classical models were retrieved with such an approach while the GBS orbits were calculated via the Monte Carlo approximation. 
These simulations show that three orbits are informative to discriminate among different gaussian samplers until the photon-number distribution is centered in the dilute regime. On the one hand, it is worth noting that the thermal light curve lies in the same plane of the GBS data but with somehow a smaller radius. The reason is that thermal radiation displays a non-zero probability to generate an odd number of photons. On the other hand, the distinguishability moves the two curves towards another plane that exhibits higher values of the probability of the orbit $[1,1,\dots 1]$. The physical intuition behind this behavior is that distinguishable particles do not interfere and, consequently, they have a lower probability of bunching. 

To prove the effectiveness of feature vectors to validate a genuine GBS device of any size, we train a classifier such as feed-forward neural network with the data reported in Fig. \ref{fig:fv_scaling}. Experimental details are provided in Appendix \ref{app:3}. Here the samples correspond to different experiment layouts with number of modes $m = n^2$, and the number of post-selected photons varying in $n \in [4, 6, \dots 22]$. The size of the collected samples was $\sim 10^5$ for the classical gaussian states that generate a fraction of $\sim 10^3-10^4$ output configurations in the orbits under investigation. For the GBS data, we performed $\sim 10^4$ Monte Carlo extractions for the orbits probability estimation. The classifier reaches high level of accuracy, greater than 99\%. We performed a further study reported in Fig. \ref{fig:fv_scaling}b to check the ability of the network to generalize 
to GBSs sizes not included in the training stage. To this aim we have trained the network with the data of Fig.~\ref{fig:fv_scaling}a up to $n = 12, 14, 16$, and subsequently computed the classification accuracy for the data with $n = 18, 20, 22$. The latter has been estimated on $100$ set of GBS for each $n$ and on $10$ independent training.

\vspace{1ex}\noindent\textbf{Validation via graph kernels.}
Other interesting quantities linked to feature vectors are the graph kernels, which can be employed to define a second method for validation. Here we study the linear kernels defined as the scalar product between pairs of feature vectors. 
This method is less demanding in terms of number of measurements since it works even in the case where only samples from a given number $n$ of photons are post-selected at the output. In Fig.~\ref{fig:kernels}a we report the distributions of kernels for feature vectors normalized to the $3$-dimensional orbits space for a given number $n$ of post-selected photons. We note that kernels from distinguishable SMVS and distinguishable thermal states  (Fig. \ref{fig:kernels}a) display the same gaussian distribution of the indistinguishable case, but they are centered at different kernel values for any $n$. Indeed, each histogram in the figure corresponds to the data of Fig. \ref{fig:leaves} for the $100$ sub-graph identified by $n=14$ and $n=16$. The coherent light data display the same average but show a larger variance. These differences highlighted in Fig.\ref{fig:kernels}a can be exploited to discriminate the coherent and distinguishable particles hypotheses. 
To do this, we only require for the optical circuit to be reconfigurable, and perform enough experiments to retrieve the kernel distributions. Note that the number of kernels scales exponentially with the number of experiments, i.e. the number of sampled different graphs. More precisely, the number of kernels after $N$ experiments is $\begin{pmatrix}
N \\
2
\end{pmatrix}$. Thus, the kernels average and variance can be retrieved in a reasonable number of measurements as investigated in Fig.~\ref{fig:kernels}b-c.
The distributions of kernels from thermal samplers (not shown in the figure) are centered at the same values of genuine GBS with the same gaussian distribution. Thus, the discrimination of data from thermal indistinguishable emitters still requires the measurement of different $n$ number of photons in the outputs. This is not surprising if we consider the distribution of the feature vectors in Fig.~\ref{fig:leaves}. They display the same dispersion of the GBS data and, since we are now considering only the space of configurations with a given number of photons, the clouds collapse on each other.

\section*{Discussion}
In this work, we have presented a new approach to GBS validation that exploits the intrinsic connection between photon counting from specific classes of gaussian states of light and counting of perfect matchings in undirected graphs. Despite GBS-based algorithms in graph theory still need further studies to clarify their actual effectiveness and advantage with respect to the classical counterparts, the tools introduced in this context turn out to be informative in the framework of GBS experiments verification. We have seen how the feature vectors together with the graph kernels extracted from photon counting indicate the quantum nature of the sampling process. In fact, these quantities are very sensitive to imperfections that could occur in actual experiments, such as photon losses and distinguishability \cite{Shuld_GBS_graphsimilarity}. These two effects drive the device to act more similarly to thermal and distinguishable particles samplers that can be simulated efficiently by classical means. 

The methods based on graph feature vectors and kernel distributions require a reasonable number of samples due to the coarse-graining of the output space of GBSs. 
The method based on graph kernels requires fewer experiments with different graphs, in turn requiring the capability to tune the optical circuit $U$ and the squeezing parameters $s_i$. Nowadays, recent experimental results on integrated reconfigurable circuits \cite{Arrazola2021, Taballione_2021, hoch2021boson} enable large tunability and dimension of the matrix $U$. In addition, squeezing parameters can be tuned by changing the power of the pump laser that generates squeezed light from nonlinear crystals, and by tuning the relative squeezing parameters phases as recently demonstrated in~\cite{zhong2021phaseprogrammable}.

Further improvements to the approach adopted in this work can be foreseen. For instance, these include exploiting a more extensive orbit set or larger coarse-graining. These modifications could help in the validation of larger-scale instances of GBS. For example, it is possible to observe from Fig. \ref{fig:leaves} and Fig. \ref{fig:fv_scaling} that the orbits probabilities tend to zero with larger size due to the increasing dimension of the GBS Hilbert space.
A future perspective of such investigation may be the extension in the regime that exploits threshold detectors. This configuration has been adopted to prove quantum advantage, but its connection with graph feature vectors has not been investigated yet. 

\section*{Acknowledgments} 
This work is supported by the ERC Advanced grant QU-BOSS (Grant Agreement No. 884676) and ERC Starting grant SPECGEO (no. 802554). The authors wish to acknowledge financial support also by MIUR (Ministero dell’Istruzione, dell’Università e della Ricerca) via project PRIN 2017 “Taming complexity via QUantum Strategies: a Hybrid Integrated Photonic approach” (QUSHIP - Id. 2017SRNBRK). N.S. acknowledges funding from Sapienza Universit\`a di Roma via Bando Ricerca 2020: Progetti di Ricerca Piccoli, project "Validation of Boson Sampling via Machine Learning". 

\appendix
\section{Sampling from gaussian states}\label{app:sampling}
\noindent
\textbf{Thermal and coherent light.} Not all gaussian states display the same sampling complexity of indistinguishable squeezed states.
For example, in the case of $m$ indistinguishable thermal light emitters, we have that the sampling matrix $A$ in Fig \ref{fig:haf_perm}a has $B=0$ and $C \ne 0$. Then, according to the relationship in \eqref{eq:permanent} the probability to detect $\vec{n}$ is
\begin{equation}
\text{Pr}(\vec{n})_{\text{Th}}=     \frac{\text{Per}\,({C}_{\vec{n}})}{\prod_{\vec{n}}{n_j}! \prod_{i=1}^m(1 + \langle n_i \rangle)},
\end{equation}
where $\langle n_i\rangle$ is the mean photon number associated to each thermal source $\rho_i$ in the input and $C_{\vec{n}}$ the sub-matrix of $C$. Sampling from this distribution is not hard as in the case of SMSV input states, although the expression requires the calculation of a permanent. Indeed, the matrix $C$ can be decomposed as $C=U \text{diag}(\tau_i, \dots, \tau_m)U^\dagger$ with $\tau_i = \langle n_i \rangle / (1 + \langle n_i \rangle)$ and is a positive semi-definite matrix. In this special case, the permanent can be approximated with classical resources \cite{perm_pos_semi}. More precisely, sampling from a thermal state is a problem in the class $\text{BPP}^{\text{NP}}$ \cite{wcqoscct}, that includes problems easier requiring lower time resources than those belonging to the \#P-complete one. An alternative way to sample from a thermal state is through the P-representation of the electromagnetic field \cite{Glauber2007, loudon1983quantum}, i.e the expression of the state as superpositions or mixtures of coherent states $\ket{\alpha}$. For classical states of light, such as thermal states, the P-function is a probability distribution, more precisely it is gaussian  $\text{P}_{\text{Th}}(\alpha_i) = \frac{1}{\pi\langle n_i \rangle} e^{-\frac{|\alpha_i|^2}{\langle n_i \rangle}}$. Then, sampling a string $\vec{n}$ can be simulated by extracting a set of $\alpha_i$ from $\text{P}_{\text{Th}}(\alpha_i)$ and considering the evolution of a coherent state in a linear optics interferometer. The latter transformation maps the state $\alpha_i$ in another coherent state $\beta_j$, according to $\beta_j = \sum_{i=1}^m U_{ij} \alpha_i$. The probability to detect the configuration $\vec{n}$ in the output, given a set of coherent states in the input, is given by:
\begin{equation}
    Pr(\vec{n})_{\text{Coh}} = \prod_{i=1}^m \frac{e^{-|\beta_i|^2} |\beta_i|^{2n_i}}{n_i!}\,.
    \label{eq:coherent}
\end{equation}
Such expression is the product of $m$ poissonian distributions and it can be sampled in polynomial time \cite{wcqoscct}. Furthermore, Eq. \eqref{eq:coherent} can be directly employed to perform classical sampling with coherent state inputs.

\noindent
\textbf{Distinguishable emitters of thermal and squeezed light.} The last possible scenario corresponds to sampling from a set of distinguishable gaussian states. The evolution can be simulated by independently sampling photons from the photon-number distribution of each gaussian source, and by considering the single-photon distribution after the interferometer. The distinguishability among photons generated from different sources permits to sample each input mode independently, without the complexity introduced by quantum interference. This is the case of distinguishable SMSV, states, and is analogous to the scenario for distinguishable thermal light emitters. Hence, efficient simulation of distinguishable gaussian states can per performed classically, thus not requiring a quantum processor.

\section{Graphs employed in the numerical simulations}
One of the main novelties in the GBS paradigm is the possibility to encode any symmetric matrix in the sampling process. This feature permits to identify relations between photon counting and the graph represented by its symmetric adjacency matrix. In the simulations presented in this work we consider the graphs resulting from $A = U \text{diag}(\tanh{s_1}, \dots, \tanh{s_m} )U^t$, where the unitary matrix $U$ of the optical circuit and the squeezing parameters $\{s_i\}$ were set as follows. The circuits $U$ were randomly generated from the Haar distribution of unitary matrices. This choice guarantees on one hand to reproduce the most general conditions, and, on the other, to exclude the existence of any symmetry inside the sampling matrix that could undermine the complexity of the problem. For what concerns the squeezing parameters, we set their values to obtain a photon number distribution centered around a given $n$. The expression of such a distribution is analyzed in details in~\cite{DetailedstudyGBS}. 

The parameter settings of the classical-simulable gaussian states were chosen to reproduce distributions as close as possible to the genuine GBS ones. To this end, we consider the evolution through the same circuit $U$. In addition, the gaussian sources were set to emit the same average number of photons of the squeezed sources. This implies setting $\langle n_i \rangle = \sinh^2 s_i$.

\section{Neural network for data classification}
\label{app:3}

In this Section we describe the experimental settings for the validation task presented in the main text. 
The dataset is made of 100 feature vectors corresponding to samples of number of photons $n$ ranging from 4 to 22 and scattered by optical circuits with $m = n^2$ optical modes (see Fig. \ref{fig:fv_scaling}) The feature vectors are labelled by two classes, distinguishing 3-dimensional features vectors extracted from genuine GBS samples, from the other samplers (see Appendix \ref{app:sampling}). To solve the validation task, we employ a Multi-Layer Perceptron (MLP) binary classifier. The architecture is composed of two stacked linear layers with ReLU activations, and a batch normalization layer \cite{Ioffe2015Feb} to improve the gradient flow in the backward pass. A final linear layer with a sigmoid activation is added to output the corresponding class prediction. The overall model is trained with a binary cross entropy loss. The MLP converged after 20 epochs, reaching an accuracy score of 99\% over the test set.
The model is implemented using PyTorch \cite{pytorch} and the code was run on a computer with a GPU NVIDIA GeForce GTX 1060 with 3GB of RAM.


\begin{thebibliography}{58}%
	\makeatletter
	\providecommand \@ifxundefined [1]{%
		\@ifx{#1\undefined}
	}%
	\providecommand \@ifnum [1]{%
		\ifnum #1\expandafter \@firstoftwo
		\else \expandafter \@secondoftwo
		\fi
	}%
	\providecommand \@ifx [1]{%
		\ifx #1\expandafter \@firstoftwo
		\else \expandafter \@secondoftwo
		\fi
	}%
	\providecommand \natexlab [1]{#1}%
	\providecommand \enquote  [1]{``#1''}%
	\providecommand \bibnamefont  [1]{#1}%
	\providecommand \bibfnamefont [1]{#1}%
	\providecommand \citenamefont [1]{#1}%
	\providecommand \href@noop [0]{\@secondoftwo}%
	\providecommand \href [0]{\begingroup \@sanitize@url \@href}%
	\providecommand \@href[1]{\@@startlink{#1}\@@href}%
	\providecommand \@@href[1]{\endgroup#1\@@endlink}%
	\providecommand \@sanitize@url [0]{\catcode `\\12\catcode `\$12\catcode
		`\&12\catcode `\#12\catcode `\^12\catcode `\_12\catcode `\%12\relax}%
	\providecommand \@@startlink[1]{}%
	\providecommand \@@endlink[0]{}%
	\providecommand \url  [0]{\begingroup\@sanitize@url \@url }%
	\providecommand \@url [1]{\endgroup\@href {#1}{\urlprefix }}%
	\providecommand \urlprefix  [0]{URL }%
	\providecommand \Eprint [0]{\href }%
	\providecommand \doibase [0]{https://doi.org/}%
	\providecommand \selectlanguage [0]{\@gobble}%
	\providecommand \bibinfo  [0]{\@secondoftwo}%
	\providecommand \bibfield  [0]{\@secondoftwo}%
	\providecommand \translation [1]{[#1]}%
	\providecommand \BibitemOpen [0]{}%
	\providecommand \bibitemStop [0]{}%
	\providecommand \bibitemNoStop [0]{.\EOS\space}%
	\providecommand \EOS [0]{\spacefactor3000\relax}%
	\providecommand \BibitemShut  [1]{\csname bibitem#1\endcsname}%
	\let\auto@bib@innerbib\@empty
	\bibitem [{\citenamefont {Harrow}\ and\ \citenamefont
		{Montanaro}(2017)}]{Harrow2017_supremacy}%
	\BibitemOpen
	\bibfield  {author} {\bibinfo {author} {\bibfnamefont {A.~W.}\ \bibnamefont
			{Harrow}}\ and\ \bibinfo {author} {\bibfnamefont {A.}~\bibnamefont
			{Montanaro}},\ }\bibfield  {title} {\bibinfo {title} {Quantum computational
			supremacy},\ }\href {https://doi.org/10.1038/nature23458} {\bibfield
		{journal} {\bibinfo  {journal} {Nature}\ }\textbf {\bibinfo {volume} {549}},\
		\bibinfo {pages} {203} (\bibinfo {year} {2017})}\BibitemShut {NoStop}%
	\bibitem [{\citenamefont {Arute}\ \emph {et~al.}(2019)\citenamefont {Arute},
		\citenamefont {Arya}, \citenamefont {Babbush}, \citenamefont {Bacon},
		\citenamefont {Bardin}, \citenamefont {Barends}, \citenamefont {Biswas},
		\citenamefont {Boixo}, \citenamefont {Brandao}, \citenamefont {Buell},
		\citenamefont {Burkett}, \citenamefont {Chen}, \citenamefont {Chen},
		\citenamefont {Chiaro}, \citenamefont {Collins}, \citenamefont {Courtney},
		\citenamefont {Dunsworth}, \citenamefont {Farhi}, \citenamefont {Foxen},
		\citenamefont {Fowler}, \citenamefont {Gidney}, \citenamefont {Giustina},
		\citenamefont {Graff}, \citenamefont {Guerin}, \citenamefont {Habegger},
		\citenamefont {Harrigan}, \citenamefont {Hartmann}, \citenamefont {Ho},
		\citenamefont {Hoffmann}, \citenamefont {Huang}, \citenamefont {Humble},
		\citenamefont {Isakov}, \citenamefont {Jeffrey}, \citenamefont {Jiang},
		\citenamefont {Kafri}, \citenamefont {Kechedzhi}, \citenamefont {Kelly},
		\citenamefont {Klimov}, \citenamefont {Knysh}, \citenamefont {Korotkov},
		\citenamefont {Kostritsa}, \citenamefont {Landhuis}, \citenamefont
		{Lindmark}, \citenamefont {Lucero}, \citenamefont {Lyakh}, \citenamefont
		{Mandr{\`a}}, \citenamefont {McClean}, \citenamefont {McEwen}, \citenamefont
		{Megrant}, \citenamefont {Mi}, \citenamefont {Michielsen}, \citenamefont
		{Mohseni}, \citenamefont {Mutus}, \citenamefont {Naaman}, \citenamefont
		{Neeley}, \citenamefont {Neill}, \citenamefont {Niu}, \citenamefont {Ostby},
		\citenamefont {Petukhov}, \citenamefont {Platt}, \citenamefont {Quintana},
		\citenamefont {Rieffel}, \citenamefont {Roushan}, \citenamefont {Rubin},
		\citenamefont {Sank}, \citenamefont {Satzinger}, \citenamefont {Smelyanskiy},
		\citenamefont {Sung}, \citenamefont {Trevithick}, \citenamefont
		{Vainsencher}, \citenamefont {Villalonga}, \citenamefont {White},
		\citenamefont {Yao}, \citenamefont {Yeh}, \citenamefont {Zalcman},
		\citenamefont {Neven},\ and\ \citenamefont {Martinis}}]{Arute2019}%
	\BibitemOpen
	\bibfield  {author} {\bibinfo {author} {\bibfnamefont {F.}~\bibnamefont
			{Arute}}, \bibinfo {author} {\bibfnamefont {K.}~\bibnamefont {Arya}},
		\bibinfo {author} {\bibfnamefont {R.}~\bibnamefont {Babbush}}, \bibinfo
		{author} {\bibfnamefont {D.}~\bibnamefont {Bacon}}, \bibinfo {author}
		{\bibfnamefont {J.~C.}\ \bibnamefont {Bardin}}, \bibinfo {author}
		{\bibfnamefont {R.}~\bibnamefont {Barends}}, \bibinfo {author} {\bibfnamefont
			{R.}~\bibnamefont {Biswas}}, \bibinfo {author} {\bibfnamefont
			{S.}~\bibnamefont {Boixo}}, \bibinfo {author} {\bibfnamefont {F.~G. S.~L.}\
			\bibnamefont {Brandao}}, \bibinfo {author} {\bibfnamefont {D.~A.}\
			\bibnamefont {Buell}}, \bibinfo {author} {\bibfnamefont {B.}~\bibnamefont
			{Burkett}}, \bibinfo {author} {\bibfnamefont {Y.}~\bibnamefont {Chen}},
		\bibinfo {author} {\bibfnamefont {Z.}~\bibnamefont {Chen}}, \bibinfo {author}
		{\bibfnamefont {B.}~\bibnamefont {Chiaro}}, \bibinfo {author} {\bibfnamefont
			{R.}~\bibnamefont {Collins}}, \bibinfo {author} {\bibfnamefont
			{W.}~\bibnamefont {Courtney}}, \bibinfo {author} {\bibfnamefont
			{A.}~\bibnamefont {Dunsworth}}, \bibinfo {author} {\bibfnamefont
			{E.}~\bibnamefont {Farhi}}, \bibinfo {author} {\bibfnamefont
			{B.}~\bibnamefont {Foxen}}, \bibinfo {author} {\bibfnamefont
			{A.}~\bibnamefont {Fowler}}, \bibinfo {author} {\bibfnamefont
			{C.}~\bibnamefont {Gidney}}, \bibinfo {author} {\bibfnamefont
			{M.}~\bibnamefont {Giustina}}, \bibinfo {author} {\bibfnamefont
			{R.}~\bibnamefont {Graff}}, \bibinfo {author} {\bibfnamefont
			{K.}~\bibnamefont {Guerin}}, \bibinfo {author} {\bibfnamefont
			{S.}~\bibnamefont {Habegger}}, \bibinfo {author} {\bibfnamefont {M.~P.}\
			\bibnamefont {Harrigan}}, \bibinfo {author} {\bibfnamefont {M.~J.}\
			\bibnamefont {Hartmann}}, \bibinfo {author} {\bibfnamefont {A.}~\bibnamefont
			{Ho}}, \bibinfo {author} {\bibfnamefont {M.}~\bibnamefont {Hoffmann}},
		\bibinfo {author} {\bibfnamefont {T.}~\bibnamefont {Huang}}, \bibinfo
		{author} {\bibfnamefont {T.~S.}\ \bibnamefont {Humble}}, \bibinfo {author}
		{\bibfnamefont {S.~V.}\ \bibnamefont {Isakov}}, \bibinfo {author}
		{\bibfnamefont {E.}~\bibnamefont {Jeffrey}}, \bibinfo {author} {\bibfnamefont
			{Z.}~\bibnamefont {Jiang}}, \bibinfo {author} {\bibfnamefont
			{D.}~\bibnamefont {Kafri}}, \bibinfo {author} {\bibfnamefont
			{K.}~\bibnamefont {Kechedzhi}}, \bibinfo {author} {\bibfnamefont
			{J.}~\bibnamefont {Kelly}}, \bibinfo {author} {\bibfnamefont {P.~V.}\
			\bibnamefont {Klimov}}, \bibinfo {author} {\bibfnamefont {S.}~\bibnamefont
			{Knysh}}, \bibinfo {author} {\bibfnamefont {A.}~\bibnamefont {Korotkov}},
		\bibinfo {author} {\bibfnamefont {F.}~\bibnamefont {Kostritsa}}, \bibinfo
		{author} {\bibfnamefont {D.}~\bibnamefont {Landhuis}}, \bibinfo {author}
		{\bibfnamefont {M.}~\bibnamefont {Lindmark}}, \bibinfo {author}
		{\bibfnamefont {E.}~\bibnamefont {Lucero}}, \bibinfo {author} {\bibfnamefont
			{D.}~\bibnamefont {Lyakh}}, \bibinfo {author} {\bibfnamefont
			{S.}~\bibnamefont {Mandr{\`a}}}, \bibinfo {author} {\bibfnamefont {J.~R.}\
			\bibnamefont {McClean}}, \bibinfo {author} {\bibfnamefont {M.}~\bibnamefont
			{McEwen}}, \bibinfo {author} {\bibfnamefont {A.}~\bibnamefont {Megrant}},
		\bibinfo {author} {\bibfnamefont {X.}~\bibnamefont {Mi}}, \bibinfo {author}
		{\bibfnamefont {K.}~\bibnamefont {Michielsen}}, \bibinfo {author}
		{\bibfnamefont {M.}~\bibnamefont {Mohseni}}, \bibinfo {author} {\bibfnamefont
			{J.}~\bibnamefont {Mutus}}, \bibinfo {author} {\bibfnamefont
			{O.}~\bibnamefont {Naaman}}, \bibinfo {author} {\bibfnamefont
			{M.}~\bibnamefont {Neeley}}, \bibinfo {author} {\bibfnamefont
			{C.}~\bibnamefont {Neill}}, \bibinfo {author} {\bibfnamefont {M.~Y.}\
			\bibnamefont {Niu}}, \bibinfo {author} {\bibfnamefont {E.}~\bibnamefont
			{Ostby}}, \bibinfo {author} {\bibfnamefont {A.}~\bibnamefont {Petukhov}},
		\bibinfo {author} {\bibfnamefont {J.~C.}\ \bibnamefont {Platt}}, \bibinfo
		{author} {\bibfnamefont {C.}~\bibnamefont {Quintana}}, \bibinfo {author}
		{\bibfnamefont {E.~G.}\ \bibnamefont {Rieffel}}, \bibinfo {author}
		{\bibfnamefont {P.}~\bibnamefont {Roushan}}, \bibinfo {author} {\bibfnamefont
			{N.~C.}\ \bibnamefont {Rubin}}, \bibinfo {author} {\bibfnamefont
			{D.}~\bibnamefont {Sank}}, \bibinfo {author} {\bibfnamefont {K.~J.}\
			\bibnamefont {Satzinger}}, \bibinfo {author} {\bibfnamefont {V.}~\bibnamefont
			{Smelyanskiy}}, \bibinfo {author} {\bibfnamefont {K.~J.}\ \bibnamefont
			{Sung}}, \bibinfo {author} {\bibfnamefont {M.~D.}\ \bibnamefont
			{Trevithick}}, \bibinfo {author} {\bibfnamefont {A.}~\bibnamefont
			{Vainsencher}}, \bibinfo {author} {\bibfnamefont {B.}~\bibnamefont
			{Villalonga}}, \bibinfo {author} {\bibfnamefont {T.}~\bibnamefont {White}},
		\bibinfo {author} {\bibfnamefont {Z.~J.}\ \bibnamefont {Yao}}, \bibinfo
		{author} {\bibfnamefont {P.}~\bibnamefont {Yeh}}, \bibinfo {author}
		{\bibfnamefont {A.}~\bibnamefont {Zalcman}}, \bibinfo {author} {\bibfnamefont
			{H.}~\bibnamefont {Neven}},\ and\ \bibinfo {author} {\bibfnamefont {J.~M.}\
			\bibnamefont {Martinis}},\ }\bibfield  {title} {\bibinfo {title} {Quantum
			supremacy using a programmable superconducting processor},\ }\href
	{https://doi.org/10.1038/s41586-019-1666-5} {\bibfield  {journal} {\bibinfo
			{journal} {Nature}\ }\textbf {\bibinfo {volume} {574}},\ \bibinfo {pages}
		{505} (\bibinfo {year} {2019})}\BibitemShut {NoStop}%
	\bibitem [{\citenamefont {Wu}\ \emph {et~al.}(2021)\citenamefont {Wu},
		\citenamefont {Bao}, \citenamefont {Cao}, \citenamefont {Chen}, \citenamefont
		{Chen}, \citenamefont {Chen}, \citenamefont {Chung}, \citenamefont {Deng},
		\citenamefont {Du}, \citenamefont {Fan}, \citenamefont {Gong}, \citenamefont
		{Guo}, \citenamefont {Guo}, \citenamefont {Guo}, \citenamefont {Han},
		\citenamefont {Hong}, \citenamefont {Huang}, \citenamefont {Huo},
		\citenamefont {Li}, \citenamefont {Li}, \citenamefont {Li}, \citenamefont
		{Li}, \citenamefont {Liang}, \citenamefont {Lin}, \citenamefont {Lin},
		\citenamefont {Qian}, \citenamefont {Qiao}, \citenamefont {Rong},
		\citenamefont {Su}, \citenamefont {Sun}, \citenamefont {Wang}, \citenamefont
		{Wang}, \citenamefont {Wu}, \citenamefont {Xu}, \citenamefont {Yan},
		\citenamefont {Yang}, \citenamefont {Yang}, \citenamefont {Ye}, \citenamefont
		{Yin}, \citenamefont {Ying}, \citenamefont {Yu}, \citenamefont {Zha},
		\citenamefont {Zhang}, \citenamefont {Zhang}, \citenamefont {Zhang},
		\citenamefont {Zhang}, \citenamefont {Zhao}, \citenamefont {Zhao},
		\citenamefont {Zhou}, \citenamefont {Zhu}, \citenamefont {Lu}, \citenamefont
		{Peng}, \citenamefont {Zhu},\ and\ \citenamefont {Pan}}]{Wu_2021_supremacy}%
	\BibitemOpen
	\bibfield  {author} {\bibinfo {author} {\bibfnamefont {Y.}~\bibnamefont
			{Wu}}, \bibinfo {author} {\bibfnamefont {W.-S.}\ \bibnamefont {Bao}},
		\bibinfo {author} {\bibfnamefont {S.}~\bibnamefont {Cao}}, \bibinfo {author}
		{\bibfnamefont {F.}~\bibnamefont {Chen}}, \bibinfo {author} {\bibfnamefont
			{M.-C.}\ \bibnamefont {Chen}}, \bibinfo {author} {\bibfnamefont
			{X.}~\bibnamefont {Chen}}, \bibinfo {author} {\bibfnamefont {T.-H.}\
			\bibnamefont {Chung}}, \bibinfo {author} {\bibfnamefont {H.}~\bibnamefont
			{Deng}}, \bibinfo {author} {\bibfnamefont {Y.}~\bibnamefont {Du}}, \bibinfo
		{author} {\bibfnamefont {D.}~\bibnamefont {Fan}}, \bibinfo {author}
		{\bibfnamefont {M.}~\bibnamefont {Gong}}, \bibinfo {author} {\bibfnamefont
			{C.}~\bibnamefont {Guo}}, \bibinfo {author} {\bibfnamefont {C.}~\bibnamefont
			{Guo}}, \bibinfo {author} {\bibfnamefont {S.}~\bibnamefont {Guo}}, \bibinfo
		{author} {\bibfnamefont {L.}~\bibnamefont {Han}}, \bibinfo {author}
		{\bibfnamefont {L.}~\bibnamefont {Hong}}, \bibinfo {author} {\bibfnamefont
			{H.-L.}\ \bibnamefont {Huang}}, \bibinfo {author} {\bibfnamefont {Y.-H.}\
			\bibnamefont {Huo}}, \bibinfo {author} {\bibfnamefont {L.}~\bibnamefont
			{Li}}, \bibinfo {author} {\bibfnamefont {N.}~\bibnamefont {Li}}, \bibinfo
		{author} {\bibfnamefont {S.}~\bibnamefont {Li}}, \bibinfo {author}
		{\bibfnamefont {Y.}~\bibnamefont {Li}}, \bibinfo {author} {\bibfnamefont
			{F.}~\bibnamefont {Liang}}, \bibinfo {author} {\bibfnamefont
			{C.}~\bibnamefont {Lin}}, \bibinfo {author} {\bibfnamefont {J.}~\bibnamefont
			{Lin}}, \bibinfo {author} {\bibfnamefont {H.}~\bibnamefont {Qian}}, \bibinfo
		{author} {\bibfnamefont {D.}~\bibnamefont {Qiao}}, \bibinfo {author}
		{\bibfnamefont {H.}~\bibnamefont {Rong}}, \bibinfo {author} {\bibfnamefont
			{H.}~\bibnamefont {Su}}, \bibinfo {author} {\bibfnamefont {L.}~\bibnamefont
			{Sun}}, \bibinfo {author} {\bibfnamefont {L.}~\bibnamefont {Wang}}, \bibinfo
		{author} {\bibfnamefont {S.}~\bibnamefont {Wang}}, \bibinfo {author}
		{\bibfnamefont {D.}~\bibnamefont {Wu}}, \bibinfo {author} {\bibfnamefont
			{Y.}~\bibnamefont {Xu}}, \bibinfo {author} {\bibfnamefont {K.}~\bibnamefont
			{Yan}}, \bibinfo {author} {\bibfnamefont {W.}~\bibnamefont {Yang}}, \bibinfo
		{author} {\bibfnamefont {Y.}~\bibnamefont {Yang}}, \bibinfo {author}
		{\bibfnamefont {Y.}~\bibnamefont {Ye}}, \bibinfo {author} {\bibfnamefont
			{J.}~\bibnamefont {Yin}}, \bibinfo {author} {\bibfnamefont {C.}~\bibnamefont
			{Ying}}, \bibinfo {author} {\bibfnamefont {J.}~\bibnamefont {Yu}}, \bibinfo
		{author} {\bibfnamefont {C.}~\bibnamefont {Zha}}, \bibinfo {author}
		{\bibfnamefont {C.}~\bibnamefont {Zhang}}, \bibinfo {author} {\bibfnamefont
			{H.}~\bibnamefont {Zhang}}, \bibinfo {author} {\bibfnamefont
			{K.}~\bibnamefont {Zhang}}, \bibinfo {author} {\bibfnamefont
			{Y.}~\bibnamefont {Zhang}}, \bibinfo {author} {\bibfnamefont
			{H.}~\bibnamefont {Zhao}}, \bibinfo {author} {\bibfnamefont {Y.}~\bibnamefont
			{Zhao}}, \bibinfo {author} {\bibfnamefont {L.}~\bibnamefont {Zhou}}, \bibinfo
		{author} {\bibfnamefont {Q.}~\bibnamefont {Zhu}}, \bibinfo {author}
		{\bibfnamefont {C.-Y.}\ \bibnamefont {Lu}}, \bibinfo {author} {\bibfnamefont
			{C.-Z.}\ \bibnamefont {Peng}}, \bibinfo {author} {\bibfnamefont
			{X.}~\bibnamefont {Zhu}},\ and\ \bibinfo {author} {\bibfnamefont {J.-W.}\
			\bibnamefont {Pan}},\ }\bibfield  {title} {\bibinfo {title} {Strong quantum
			computational advantage using a superconducting quantum processor},\ }\href
	{https://doi.org/10.1103/PhysRevLett.127.180501} {\bibfield  {journal}
		{\bibinfo  {journal} {Phys. Rev. Lett.}\ }\textbf {\bibinfo {volume} {127}},\
		\bibinfo {pages} {180501} (\bibinfo {year} {2021})}\BibitemShut {NoStop}%
	\bibitem [{\citenamefont {Zhong}\ \emph {et~al.}(2020)\citenamefont {Zhong},
		\citenamefont {Wang}, \citenamefont {Deng}, \citenamefont {Chen},
		\citenamefont {Peng}, \citenamefont {Luo}, \citenamefont {Qin}, \citenamefont
		{Wu}, \citenamefont {Ding}, \citenamefont {Hu}, \citenamefont {Hu},
		\citenamefont {Yang}, \citenamefont {Zhang}, \citenamefont {Li},
		\citenamefont {Li}, \citenamefont {Jiang}, \citenamefont {Gan}, \citenamefont
		{Yang}, \citenamefont {You}, \citenamefont {Wang}, \citenamefont {Li},
		\citenamefont {Liu}, \citenamefont {Lu},\ and\ \citenamefont
		{Pan}}]{Zhong_GBS_supremacy}%
	\BibitemOpen
	\bibfield  {author} {\bibinfo {author} {\bibfnamefont {H.-S.}\ \bibnamefont
			{Zhong}}, \bibinfo {author} {\bibfnamefont {H.}~\bibnamefont {Wang}},
		\bibinfo {author} {\bibfnamefont {Y.-H.}\ \bibnamefont {Deng}}, \bibinfo
		{author} {\bibfnamefont {M.-C.}\ \bibnamefont {Chen}}, \bibinfo {author}
		{\bibfnamefont {L.-C.}\ \bibnamefont {Peng}}, \bibinfo {author}
		{\bibfnamefont {Y.-H.}\ \bibnamefont {Luo}}, \bibinfo {author} {\bibfnamefont
			{J.}~\bibnamefont {Qin}}, \bibinfo {author} {\bibfnamefont {D.}~\bibnamefont
			{Wu}}, \bibinfo {author} {\bibfnamefont {X.}~\bibnamefont {Ding}}, \bibinfo
		{author} {\bibfnamefont {Y.}~\bibnamefont {Hu}}, \bibinfo {author}
		{\bibfnamefont {P.}~\bibnamefont {Hu}}, \bibinfo {author} {\bibfnamefont
			{X.-Y.}\ \bibnamefont {Yang}}, \bibinfo {author} {\bibfnamefont {W.-J.}\
			\bibnamefont {Zhang}}, \bibinfo {author} {\bibfnamefont {H.}~\bibnamefont
			{Li}}, \bibinfo {author} {\bibfnamefont {Y.}~\bibnamefont {Li}}, \bibinfo
		{author} {\bibfnamefont {X.}~\bibnamefont {Jiang}}, \bibinfo {author}
		{\bibfnamefont {L.}~\bibnamefont {Gan}}, \bibinfo {author} {\bibfnamefont
			{G.}~\bibnamefont {Yang}}, \bibinfo {author} {\bibfnamefont {L.}~\bibnamefont
			{You}}, \bibinfo {author} {\bibfnamefont {Z.}~\bibnamefont {Wang}}, \bibinfo
		{author} {\bibfnamefont {L.}~\bibnamefont {Li}}, \bibinfo {author}
		{\bibfnamefont {N.-L.}\ \bibnamefont {Liu}}, \bibinfo {author} {\bibfnamefont
			{C.-Y.}\ \bibnamefont {Lu}},\ and\ \bibinfo {author} {\bibfnamefont {J.-W.}\
			\bibnamefont {Pan}},\ }\bibfield  {title} {\bibinfo {title} {Quantum
			computational advantage using photons},\ }\href
	{https://doi.org/10.1126/science.abe8770} {\bibfield  {journal} {\bibinfo
			{journal} {Science}\ }\textbf {\bibinfo {volume} {370}},\ \bibinfo {pages}
		{1460} (\bibinfo {year} {2020})}\BibitemShut {NoStop}%
	\bibitem [{\citenamefont {Zhong}\ \emph {et~al.}(2021)\citenamefont {Zhong},
		\citenamefont {Deng}, \citenamefont {Qin}, \citenamefont {Wang},
		\citenamefont {Chen}, \citenamefont {Peng}, \citenamefont {Luo},
		\citenamefont {Wu}, \citenamefont {Gong}, \citenamefont {Su}, \citenamefont
		{Hu}, \citenamefont {Hu}, \citenamefont {Yang}, \citenamefont {Zhang},
		\citenamefont {Li}, \citenamefont {Li}, \citenamefont {Jiang}, \citenamefont
		{Gan}, \citenamefont {Yang}, \citenamefont {You}, \citenamefont {Wang},
		\citenamefont {Li}, \citenamefont {Liu}, \citenamefont {Renema},
		\citenamefont {Lu},\ and\ \citenamefont {Pan}}]{zhong2021phaseprogrammable}%
	\BibitemOpen
	\bibfield  {author} {\bibinfo {author} {\bibfnamefont {H.-S.}\ \bibnamefont
			{Zhong}}, \bibinfo {author} {\bibfnamefont {Y.-H.}\ \bibnamefont {Deng}},
		\bibinfo {author} {\bibfnamefont {J.}~\bibnamefont {Qin}}, \bibinfo {author}
		{\bibfnamefont {H.}~\bibnamefont {Wang}}, \bibinfo {author} {\bibfnamefont
			{M.-C.}\ \bibnamefont {Chen}}, \bibinfo {author} {\bibfnamefont {L.-C.}\
			\bibnamefont {Peng}}, \bibinfo {author} {\bibfnamefont {Y.-H.}\ \bibnamefont
			{Luo}}, \bibinfo {author} {\bibfnamefont {D.}~\bibnamefont {Wu}}, \bibinfo
		{author} {\bibfnamefont {S.-Q.}\ \bibnamefont {Gong}}, \bibinfo {author}
		{\bibfnamefont {H.}~\bibnamefont {Su}}, \bibinfo {author} {\bibfnamefont
			{Y.}~\bibnamefont {Hu}}, \bibinfo {author} {\bibfnamefont {P.}~\bibnamefont
			{Hu}}, \bibinfo {author} {\bibfnamefont {X.-Y.}\ \bibnamefont {Yang}},
		\bibinfo {author} {\bibfnamefont {W.-J.}\ \bibnamefont {Zhang}}, \bibinfo
		{author} {\bibfnamefont {H.}~\bibnamefont {Li}}, \bibinfo {author}
		{\bibfnamefont {Y.}~\bibnamefont {Li}}, \bibinfo {author} {\bibfnamefont
			{X.}~\bibnamefont {Jiang}}, \bibinfo {author} {\bibfnamefont
			{L.}~\bibnamefont {Gan}}, \bibinfo {author} {\bibfnamefont {G.}~\bibnamefont
			{Yang}}, \bibinfo {author} {\bibfnamefont {L.}~\bibnamefont {You}}, \bibinfo
		{author} {\bibfnamefont {Z.}~\bibnamefont {Wang}}, \bibinfo {author}
		{\bibfnamefont {L.}~\bibnamefont {Li}}, \bibinfo {author} {\bibfnamefont
			{N.-L.}\ \bibnamefont {Liu}}, \bibinfo {author} {\bibfnamefont {J.~J.}\
			\bibnamefont {Renema}}, \bibinfo {author} {\bibfnamefont {C.-Y.}\
			\bibnamefont {Lu}},\ and\ \bibinfo {author} {\bibfnamefont {J.-W.}\
			\bibnamefont {Pan}},\ }\bibfield  {title} {\bibinfo {title}
		{Phase-programmable gaussian boson sampling using stimulated squeezed
			light},\ }\href {https://doi.org/10.1103/PhysRevLett.127.180502} {\bibfield
		{journal} {\bibinfo  {journal} {Phys. Rev. Lett.}\ }\textbf {\bibinfo
			{volume} {127}},\ \bibinfo {pages} {180502} (\bibinfo {year}
		{2021})}\BibitemShut {NoStop}%
	\bibitem [{\citenamefont {Aaronson}\ and\ \citenamefont {Arkhipov}(2011)}]{AA}%
	\BibitemOpen
	\bibfield  {author} {\bibinfo {author} {\bibfnamefont {S.}~\bibnamefont
			{Aaronson}}\ and\ \bibinfo {author} {\bibfnamefont {A.}~\bibnamefont
			{Arkhipov}},\ }\bibfield  {title} {\bibinfo {title} {The computational
			complexity of linear optics},\ }in\ \href
	{https://doi.org/10.1145/1993636.1993682} {\emph {\bibinfo {booktitle}
			{Proceedings of the 43rd annual ACM symposium on Theory of Computing}}},\
	\bibinfo {editor} {edited by\ \bibinfo {editor} {\bibfnamefont
			{A.}~\bibnamefont {Press}}}\ (\bibinfo {year} {2011})\ pp.\ \bibinfo {pages}
	{333--342}\BibitemShut {NoStop}%
	\bibitem [{\citenamefont {Brod}\ \emph {et~al.}(2019)\citenamefont {Brod},
		\citenamefont {Galvão}, \citenamefont {Crespi}, \citenamefont {Osellame},
		\citenamefont {Spagnolo},\ and\ \citenamefont {Sciarrino}}]{Brod19review}%
	\BibitemOpen
	\bibfield  {author} {\bibinfo {author} {\bibfnamefont {D.~J.}\ \bibnamefont
			{Brod}}, \bibinfo {author} {\bibfnamefont {E.~F.}\ \bibnamefont {Galvão}},
		\bibinfo {author} {\bibfnamefont {A.}~\bibnamefont {Crespi}}, \bibinfo
		{author} {\bibfnamefont {R.}~\bibnamefont {Osellame}}, \bibinfo {author}
		{\bibfnamefont {N.}~\bibnamefont {Spagnolo}},\ and\ \bibinfo {author}
		{\bibfnamefont {F.}~\bibnamefont {Sciarrino}},\ }\bibfield  {title} {\bibinfo
		{title} {Photonic implementation of boson sampling: a review},\ }\href
	{https://doi.org/10.1117/1.AP.1.3.034001} {\bibfield  {journal} {\bibinfo
			{journal} {Advanced Photonics}\ }\textbf {\bibinfo {volume} {1}},\ \bibinfo
		{pages} {1 } (\bibinfo {year} {2019})}\BibitemShut {NoStop}%
	\bibitem [{\citenamefont {Lund}\ \emph {et~al.}(2014)\citenamefont {Lund},
		\citenamefont {Laing}, \citenamefont {Rahimi-Keshari}, \citenamefont
		{Rudolph}, \citenamefont {O'Brien},\ and\ \citenamefont {Ralph}}]{Lund_SBS}%
	\BibitemOpen
	\bibfield  {author} {\bibinfo {author} {\bibfnamefont {A.~P.}\ \bibnamefont
			{Lund}}, \bibinfo {author} {\bibfnamefont {A.}~\bibnamefont {Laing}},
		\bibinfo {author} {\bibfnamefont {S.}~\bibnamefont {Rahimi-Keshari}},
		\bibinfo {author} {\bibfnamefont {T.}~\bibnamefont {Rudolph}}, \bibinfo
		{author} {\bibfnamefont {J.~L.}\ \bibnamefont {O'Brien}},\ and\ \bibinfo
		{author} {\bibfnamefont {T.~C.}\ \bibnamefont {Ralph}},\ }\bibfield  {title}
	{\bibinfo {title} {Boson sampling from a gaussian state},\ }\href
	{https://doi.org/10.1103/PhysRevLett.113.100502} {\bibfield  {journal}
		{\bibinfo  {journal} {Phys. Rev. Lett.}\ }\textbf {\bibinfo {volume} {113}},\
		\bibinfo {pages} {100502} (\bibinfo {year} {2014})}\BibitemShut {NoStop}%
	\bibitem [{\citenamefont {Rahimi-Keshari}\ \emph {et~al.}(2015)\citenamefont
		{Rahimi-Keshari}, \citenamefont {Lund},\ and\ \citenamefont
		{Ralph}}]{wcqoscct}%
	\BibitemOpen
	\bibfield  {author} {\bibinfo {author} {\bibfnamefont {S.}~\bibnamefont
			{Rahimi-Keshari}}, \bibinfo {author} {\bibfnamefont {A.~P.}\ \bibnamefont
			{Lund}},\ and\ \bibinfo {author} {\bibfnamefont {T.~C.}\ \bibnamefont
			{Ralph}},\ }\bibfield  {title} {\bibinfo {title} {What can quantum optics say
			about computational complexity theory?},\ }\href
	{https://doi.org/10.1103/PhysRevLett.114.060501} {\bibfield  {journal}
		{\bibinfo  {journal} {Phys. Rev. Lett.}\ }\textbf {\bibinfo {volume} {114}},\
		\bibinfo {pages} {060501} (\bibinfo {year} {2015})}\BibitemShut {NoStop}%
	\bibitem [{\citenamefont {Hamilton}\ \emph {et~al.}(2017)\citenamefont
		{Hamilton}, \citenamefont {Kruse}, \citenamefont {Sansoni}, \citenamefont
		{Barkhofen}, \citenamefont {Silberhorn},\ and\ \citenamefont
		{Jex}}]{Hamilton2017}%
	\BibitemOpen
	\bibfield  {author} {\bibinfo {author} {\bibfnamefont {C.~S.}\ \bibnamefont
			{Hamilton}}, \bibinfo {author} {\bibfnamefont {R.}~\bibnamefont {Kruse}},
		\bibinfo {author} {\bibfnamefont {L.}~\bibnamefont {Sansoni}}, \bibinfo
		{author} {\bibfnamefont {S.}~\bibnamefont {Barkhofen}}, \bibinfo {author}
		{\bibfnamefont {C.}~\bibnamefont {Silberhorn}},\ and\ \bibinfo {author}
		{\bibfnamefont {I.}~\bibnamefont {Jex}},\ }\bibfield  {title} {\bibinfo
		{title} {Gaussian boson sampling},\ }\href
	{https://doi.org/10.1103/PhysRevLett.119.170501} {\bibfield  {journal}
		{\bibinfo  {journal} {Phys. Rev. Lett.}\ }\textbf {\bibinfo {volume} {119}},\
		\bibinfo {pages} {170501} (\bibinfo {year} {2017})}\BibitemShut {NoStop}%
	\bibitem [{\citenamefont {Deshpande}\ \emph {et~al.}(2022)\citenamefont
		{Deshpande}, \citenamefont {Mehta}, \citenamefont {Vincent}, \citenamefont
		{Quesada}, \citenamefont {Hinsche}, \citenamefont {Ioannou}, \citenamefont
		{Madsen}, \citenamefont {Lavoie}, \citenamefont {Qi}, \citenamefont {Eisert},
		\citenamefont {Hangleiter}, \citenamefont {Fefferman},\ and\ \citenamefont
		{Dhand}}]{Deshpande_GBS_th_supremacy}%
	\BibitemOpen
	\bibfield  {author} {\bibinfo {author} {\bibfnamefont {A.}~\bibnamefont
			{Deshpande}}, \bibinfo {author} {\bibfnamefont {A.}~\bibnamefont {Mehta}},
		\bibinfo {author} {\bibfnamefont {T.}~\bibnamefont {Vincent}}, \bibinfo
		{author} {\bibfnamefont {N.}~\bibnamefont {Quesada}}, \bibinfo {author}
		{\bibfnamefont {M.}~\bibnamefont {Hinsche}}, \bibinfo {author} {\bibfnamefont
			{M.}~\bibnamefont {Ioannou}}, \bibinfo {author} {\bibfnamefont
			{L.}~\bibnamefont {Madsen}}, \bibinfo {author} {\bibfnamefont
			{J.}~\bibnamefont {Lavoie}}, \bibinfo {author} {\bibfnamefont
			{H.}~\bibnamefont {Qi}}, \bibinfo {author} {\bibfnamefont {J.}~\bibnamefont
			{Eisert}}, \bibinfo {author} {\bibfnamefont {D.}~\bibnamefont {Hangleiter}},
		\bibinfo {author} {\bibfnamefont {B.}~\bibnamefont {Fefferman}},\ and\
		\bibinfo {author} {\bibfnamefont {I.}~\bibnamefont {Dhand}},\ }\bibfield
	{title} {\bibinfo {title} {Quantum computational advantage via
			high-dimensional gaussian boson sampling},\ }\href
	{https://doi.org/10.1126/sciadv.abi7894} {\bibfield  {journal} {\bibinfo
			{journal} {Science Advances}\ }\textbf {\bibinfo {volume} {8}},\ \bibinfo
		{pages} {eabi7894} (\bibinfo {year} {2022})}\BibitemShut {NoStop}%
	\bibitem [{\citenamefont {Zhong}\ \emph {et~al.}(2019)\citenamefont {Zhong},
		\citenamefont {Peng}, \citenamefont {Li}, \citenamefont {Hu}, \citenamefont
		{Li}, \citenamefont {Qin}, \citenamefont {Wu}, \citenamefont {Zhang},
		\citenamefont {Li}, \citenamefont {Zhang}, \citenamefont {Wang},
		\citenamefont {You}, \citenamefont {Jiang}, \citenamefont {Li}, \citenamefont
		{Liu}, \citenamefont {Dowling}, \citenamefont {Lu},\ and\ \citenamefont
		{Pan}}]{Zhong19}%
	\BibitemOpen
	\bibfield  {author} {\bibinfo {author} {\bibfnamefont {H.-S.}\ \bibnamefont
			{Zhong}}, \bibinfo {author} {\bibfnamefont {L.-C.}\ \bibnamefont {Peng}},
		\bibinfo {author} {\bibfnamefont {Y.}~\bibnamefont {Li}}, \bibinfo {author}
		{\bibfnamefont {Y.}~\bibnamefont {Hu}}, \bibinfo {author} {\bibfnamefont
			{W.}~\bibnamefont {Li}}, \bibinfo {author} {\bibfnamefont {J.}~\bibnamefont
			{Qin}}, \bibinfo {author} {\bibfnamefont {D.}~\bibnamefont {Wu}}, \bibinfo
		{author} {\bibfnamefont {W.}~\bibnamefont {Zhang}}, \bibinfo {author}
		{\bibfnamefont {H.}~\bibnamefont {Li}}, \bibinfo {author} {\bibfnamefont
			{L.}~\bibnamefont {Zhang}}, \bibinfo {author} {\bibfnamefont
			{Z.}~\bibnamefont {Wang}}, \bibinfo {author} {\bibfnamefont {L.}~\bibnamefont
			{You}}, \bibinfo {author} {\bibfnamefont {X.}~\bibnamefont {Jiang}}, \bibinfo
		{author} {\bibfnamefont {L.}~\bibnamefont {Li}}, \bibinfo {author}
		{\bibfnamefont {N.-L.}\ \bibnamefont {Liu}}, \bibinfo {author} {\bibfnamefont
			{J.~P.}\ \bibnamefont {Dowling}}, \bibinfo {author} {\bibfnamefont {C.-Y.}\
			\bibnamefont {Lu}},\ and\ \bibinfo {author} {\bibfnamefont {J.-W.}\
			\bibnamefont {Pan}},\ }\bibfield  {title} {\bibinfo {title} {Experimental
			gaussian boson sampling},\ }\href
	{https://doi.org/https://doi.org/10.1016/j.scib.2019.04.007} {\bibfield
		{journal} {\bibinfo  {journal} {Science Bulletin}\ }\textbf {\bibinfo
			{volume} {64}},\ \bibinfo {pages} {511} (\bibinfo {year} {2019})}\BibitemShut
	{NoStop}%
	\bibitem [{\citenamefont {Paesani}\ \emph {et~al.}(2019)\citenamefont
		{Paesani}, \citenamefont {Ding}, \citenamefont {Santagati}, \citenamefont
		{Chakhmakhchyan}, \citenamefont {Vigliar}, \citenamefont {Rottwitt},
		\citenamefont {Oxenl{\o}we}, \citenamefont {Wang}, \citenamefont {Thompson},\
		and\ \citenamefont {Laing}}]{Paesani2019}%
	\BibitemOpen
	\bibfield  {author} {\bibinfo {author} {\bibfnamefont {S.}~\bibnamefont
			{Paesani}}, \bibinfo {author} {\bibfnamefont {Y.}~\bibnamefont {Ding}},
		\bibinfo {author} {\bibfnamefont {R.}~\bibnamefont {Santagati}}, \bibinfo
		{author} {\bibfnamefont {L.}~\bibnamefont {Chakhmakhchyan}}, \bibinfo
		{author} {\bibfnamefont {C.}~\bibnamefont {Vigliar}}, \bibinfo {author}
		{\bibfnamefont {K.}~\bibnamefont {Rottwitt}}, \bibinfo {author}
		{\bibfnamefont {L.~K.}\ \bibnamefont {Oxenl{\o}we}}, \bibinfo {author}
		{\bibfnamefont {J.}~\bibnamefont {Wang}}, \bibinfo {author} {\bibfnamefont
			{M.~G.}\ \bibnamefont {Thompson}},\ and\ \bibinfo {author} {\bibfnamefont
			{A.}~\bibnamefont {Laing}},\ }\bibfield  {title} {\bibinfo {title}
		{Generation and sampling of quantum states of light in a silicon chip},\
	}\href {https://doi.org/10.1038/s41567-019-0567-8} {\bibfield  {journal}
		{\bibinfo  {journal} {Nature Physics}\ }\textbf {\bibinfo {volume} {15}},\
		\bibinfo {pages} {925} (\bibinfo {year} {2019})}\BibitemShut {NoStop}%
	\bibitem [{\citenamefont {Thekkadath}\ \emph {et~al.}(2022)\citenamefont
		{Thekkadath}, \citenamefont {Sempere-Llagostera}, \citenamefont {Bell},
		\citenamefont {Patel}, \citenamefont {Kim},\ and\ \citenamefont
		{Walmsley}}]{thekkadath2022experimental}%
	\BibitemOpen
	\bibfield  {author} {\bibinfo {author} {\bibfnamefont {G.~S.}\ \bibnamefont
			{Thekkadath}}, \bibinfo {author} {\bibfnamefont {S.}~\bibnamefont
			{Sempere-Llagostera}}, \bibinfo {author} {\bibfnamefont {B.~A.}\ \bibnamefont
			{Bell}}, \bibinfo {author} {\bibfnamefont {R.~B.}\ \bibnamefont {Patel}},
		\bibinfo {author} {\bibfnamefont {M.~S.}\ \bibnamefont {Kim}},\ and\ \bibinfo
		{author} {\bibfnamefont {I.~A.}\ \bibnamefont {Walmsley}},\ }\href@noop {}
	{\bibinfo {title} {Experimental demonstration of gaussian boson sampling with
			displacement}} (\bibinfo {year} {2022}),\ \Eprint
	{https://arxiv.org/abs/2202.00634} {arXiv:2202.00634 [quant-ph]} \BibitemShut
	{NoStop}%
	\bibitem [{\citenamefont {Arrazola}\ and\ \citenamefont
		{Bromley}(2018)}]{Arrazzola_densesubgraph}%
	\BibitemOpen
	\bibfield  {author} {\bibinfo {author} {\bibfnamefont {J.~M.}\ \bibnamefont
			{Arrazola}}\ and\ \bibinfo {author} {\bibfnamefont {T.~R.}\ \bibnamefont
			{Bromley}},\ }\bibfield  {title} {\bibinfo {title} {Using gaussian boson
			sampling to find dense subgraphs},\ }\href
	{https://doi.org/10.1103/PhysRevLett.121.030503} {\bibfield  {journal}
		{\bibinfo  {journal} {Phys. Rev. Lett.}\ }\textbf {\bibinfo {volume} {121}},\
		\bibinfo {pages} {030503} (\bibinfo {year} {2018})}\BibitemShut {NoStop}%
	\bibitem [{\citenamefont {Schuld}\ \emph {et~al.}(2020)\citenamefont {Schuld},
		\citenamefont {Br\'adler}, \citenamefont {Israel}, \citenamefont {Su},\ and\
		\citenamefont {Gupt}}]{Shuld_GBS_graphsimilarity}%
	\BibitemOpen
	\bibfield  {author} {\bibinfo {author} {\bibfnamefont {M.}~\bibnamefont
			{Schuld}}, \bibinfo {author} {\bibfnamefont {K.}~\bibnamefont {Br\'adler}},
		\bibinfo {author} {\bibfnamefont {R.}~\bibnamefont {Israel}}, \bibinfo
		{author} {\bibfnamefont {D.}~\bibnamefont {Su}},\ and\ \bibinfo {author}
		{\bibfnamefont {B.}~\bibnamefont {Gupt}},\ }\bibfield  {title} {\bibinfo
		{title} {Measuring the similarity of graphs with a gaussian boson sampler},\
	}\href {https://doi.org/10.1103/PhysRevA.101.032314} {\bibfield  {journal}
		{\bibinfo  {journal} {Phys. Rev. A}\ }\textbf {\bibinfo {volume} {101}},\
		\bibinfo {pages} {032314} (\bibinfo {year} {2020})}\BibitemShut {NoStop}%
	\bibitem [{\citenamefont {Brádler}\ \emph {et~al.}(2021)\citenamefont
		{Brádler}, \citenamefont {Friedland}, \citenamefont {Izaac}, \citenamefont
		{Killoran},\ and\ \citenamefont {Su}}]{Bradler_2021}%
	\BibitemOpen
	\bibfield  {author} {\bibinfo {author} {\bibfnamefont {K.}~\bibnamefont
			{Brádler}}, \bibinfo {author} {\bibfnamefont {S.}~\bibnamefont {Friedland}},
		\bibinfo {author} {\bibfnamefont {J.}~\bibnamefont {Izaac}}, \bibinfo
		{author} {\bibfnamefont {N.}~\bibnamefont {Killoran}},\ and\ \bibinfo
		{author} {\bibfnamefont {D.}~\bibnamefont {Su}},\ }\bibfield  {title}
	{\bibinfo {title} {Graph isomorphism and gaussian boson sampling},\ }\href
	{https://doi.org/doi:10.1515/spma-2020-0132} {\bibfield  {journal} {\bibinfo
			{journal} {Special Matrices}\ }\textbf {\bibinfo {volume} {9}},\ \bibinfo
		{pages} {166} (\bibinfo {year} {2021})}\BibitemShut {NoStop}%
	\bibitem [{\citenamefont {Arrazola}\ \emph {et~al.}(2021)\citenamefont
		{Arrazola}, \citenamefont {Bergholm}, \citenamefont {Br{\'a}dler},
		\citenamefont {Bromley}, \citenamefont {Collins}, \citenamefont {Dhand},
		\citenamefont {Fumagalli}, \citenamefont {Gerrits}, \citenamefont {Goussev},
		\citenamefont {Helt}, \citenamefont {Hundal}, \citenamefont {Isacsson},
		\citenamefont {Israel}, \citenamefont {Izaac}, \citenamefont {Jahangiri},
		\citenamefont {Janik}, \citenamefont {Killoran}, \citenamefont {Kumar},
		\citenamefont {Lavoie}, \citenamefont {Lita}, \citenamefont {Mahler},
		\citenamefont {Menotti}, \citenamefont {Morrison}, \citenamefont {Nam},
		\citenamefont {Neuhaus}, \citenamefont {Qi}, \citenamefont {Quesada},
		\citenamefont {Repingon}, \citenamefont {Sabapathy}, \citenamefont {Schuld},
		\citenamefont {Su}, \citenamefont {Swinarton}, \citenamefont {Sz{\'a}va},
		\citenamefont {Tan}, \citenamefont {Tan}, \citenamefont {Vaidya},
		\citenamefont {Vernon}, \citenamefont {Zabaneh},\ and\ \citenamefont
		{Zhang}}]{Arrazola2021}%
	\BibitemOpen
	\bibfield  {author} {\bibinfo {author} {\bibfnamefont {J.~M.}\ \bibnamefont
			{Arrazola}}, \bibinfo {author} {\bibfnamefont {V.}~\bibnamefont {Bergholm}},
		\bibinfo {author} {\bibfnamefont {K.}~\bibnamefont {Br{\'a}dler}}, \bibinfo
		{author} {\bibfnamefont {T.~R.}\ \bibnamefont {Bromley}}, \bibinfo {author}
		{\bibfnamefont {M.~J.}\ \bibnamefont {Collins}}, \bibinfo {author}
		{\bibfnamefont {I.}~\bibnamefont {Dhand}}, \bibinfo {author} {\bibfnamefont
			{A.}~\bibnamefont {Fumagalli}}, \bibinfo {author} {\bibfnamefont
			{T.}~\bibnamefont {Gerrits}}, \bibinfo {author} {\bibfnamefont
			{A.}~\bibnamefont {Goussev}}, \bibinfo {author} {\bibfnamefont {L.~G.}\
			\bibnamefont {Helt}}, \bibinfo {author} {\bibfnamefont {J.}~\bibnamefont
			{Hundal}}, \bibinfo {author} {\bibfnamefont {T.}~\bibnamefont {Isacsson}},
		\bibinfo {author} {\bibfnamefont {R.~B.}\ \bibnamefont {Israel}}, \bibinfo
		{author} {\bibfnamefont {J.}~\bibnamefont {Izaac}}, \bibinfo {author}
		{\bibfnamefont {S.}~\bibnamefont {Jahangiri}}, \bibinfo {author}
		{\bibfnamefont {R.}~\bibnamefont {Janik}}, \bibinfo {author} {\bibfnamefont
			{N.}~\bibnamefont {Killoran}}, \bibinfo {author} {\bibfnamefont {S.~P.}\
			\bibnamefont {Kumar}}, \bibinfo {author} {\bibfnamefont {J.}~\bibnamefont
			{Lavoie}}, \bibinfo {author} {\bibfnamefont {A.~E.}\ \bibnamefont {Lita}},
		\bibinfo {author} {\bibfnamefont {D.~H.}\ \bibnamefont {Mahler}}, \bibinfo
		{author} {\bibfnamefont {M.}~\bibnamefont {Menotti}}, \bibinfo {author}
		{\bibfnamefont {B.}~\bibnamefont {Morrison}}, \bibinfo {author}
		{\bibfnamefont {S.~W.}\ \bibnamefont {Nam}}, \bibinfo {author} {\bibfnamefont
			{L.}~\bibnamefont {Neuhaus}}, \bibinfo {author} {\bibfnamefont {H.~Y.}\
			\bibnamefont {Qi}}, \bibinfo {author} {\bibfnamefont {N.}~\bibnamefont
			{Quesada}}, \bibinfo {author} {\bibfnamefont {A.}~\bibnamefont {Repingon}},
		\bibinfo {author} {\bibfnamefont {K.~K.}\ \bibnamefont {Sabapathy}}, \bibinfo
		{author} {\bibfnamefont {M.}~\bibnamefont {Schuld}}, \bibinfo {author}
		{\bibfnamefont {D.}~\bibnamefont {Su}}, \bibinfo {author} {\bibfnamefont
			{J.}~\bibnamefont {Swinarton}}, \bibinfo {author} {\bibfnamefont
			{A.}~\bibnamefont {Sz{\'a}va}}, \bibinfo {author} {\bibfnamefont
			{K.}~\bibnamefont {Tan}}, \bibinfo {author} {\bibfnamefont {P.}~\bibnamefont
			{Tan}}, \bibinfo {author} {\bibfnamefont {V.~D.}\ \bibnamefont {Vaidya}},
		\bibinfo {author} {\bibfnamefont {Z.}~\bibnamefont {Vernon}}, \bibinfo
		{author} {\bibfnamefont {Z.}~\bibnamefont {Zabaneh}},\ and\ \bibinfo {author}
		{\bibfnamefont {Y.}~\bibnamefont {Zhang}},\ }\bibfield  {title} {\bibinfo
		{title} {Quantum circuits with many photons on a programmable nanophotonic
			chip},\ }\href {https://doi.org/10.1038/s41586-021-03202-1} {\bibfield
		{journal} {\bibinfo  {journal} {Nature}\ }\textbf {\bibinfo {volume} {591}},\
		\bibinfo {pages} {54} (\bibinfo {year} {2021})}\BibitemShut {NoStop}%
	\bibitem [{\citenamefont {Clifford}\ and\ \citenamefont
		{Clifford}(2017)}]{clifford2017classical}%
	\BibitemOpen
	\bibfield  {author} {\bibinfo {author} {\bibfnamefont {P.}~\bibnamefont
			{Clifford}}\ and\ \bibinfo {author} {\bibfnamefont {R.}~\bibnamefont
			{Clifford}},\ }\href@noop {} {\bibinfo {title} {The classical complexity of
			boson sampling}} (\bibinfo {year} {2017}),\ \Eprint
	{https://arxiv.org/abs/1706.01260} {arXiv:1706.01260 [cs.DS]} \BibitemShut
	{NoStop}%
	\bibitem [{\citenamefont {Neville}\ \emph {et~al.}(2017)\citenamefont
		{Neville}, \citenamefont {Sparrow}, \citenamefont {Clifford}, \citenamefont
		{Johnston}, \citenamefont {Birchall}, \citenamefont {Montanaro},\ and\
		\citenamefont {Laing}}]{Neville2017}%
	\BibitemOpen
	\bibfield  {author} {\bibinfo {author} {\bibfnamefont {A.}~\bibnamefont
			{Neville}}, \bibinfo {author} {\bibfnamefont {C.}~\bibnamefont {Sparrow}},
		\bibinfo {author} {\bibfnamefont {R.}~\bibnamefont {Clifford}}, \bibinfo
		{author} {\bibfnamefont {E.}~\bibnamefont {Johnston}}, \bibinfo {author}
		{\bibfnamefont {P.~M.}\ \bibnamefont {Birchall}}, \bibinfo {author}
		{\bibfnamefont {A.}~\bibnamefont {Montanaro}},\ and\ \bibinfo {author}
		{\bibfnamefont {A.}~\bibnamefont {Laing}},\ }\bibfield  {title} {\bibinfo
		{title} {Classical boson sampling algorithms with superior performance to
			near-term experiments},\ }\href {https://doi.org/10.1038/nphys4270}
	{\bibfield  {journal} {\bibinfo  {journal} {Nature Physics}\ }\textbf
		{\bibinfo {volume} {13}},\ \bibinfo {pages} {1153 EP } (\bibinfo {year}
		{2017})}\BibitemShut {NoStop}%
	\bibitem [{\citenamefont {Quesada}\ and\ \citenamefont
		{Arrazola}(2020)}]{Quesada_exact_simulation}%
	\BibitemOpen
	\bibfield  {author} {\bibinfo {author} {\bibfnamefont {N.}~\bibnamefont
			{Quesada}}\ and\ \bibinfo {author} {\bibfnamefont {J.~M.}\ \bibnamefont
			{Arrazola}},\ }\bibfield  {title} {\bibinfo {title} {Exact simulation of
			gaussian boson sampling in polynomial space and exponential time},\ }\href
	{https://doi.org/10.1103/PhysRevResearch.2.023005} {\bibfield  {journal}
		{\bibinfo  {journal} {Phys. Rev. Research}\ }\textbf {\bibinfo {volume}
			{2}},\ \bibinfo {pages} {023005} (\bibinfo {year} {2020})}\BibitemShut
	{NoStop}%
	\bibitem [{\citenamefont {Quesada}\ \emph {et~al.}(2022)\citenamefont
		{Quesada}, \citenamefont {Chadwick}, \citenamefont {Bell}, \citenamefont
		{Arrazola}, \citenamefont {Vincent}, \citenamefont {Qi},\ and\ \citenamefont
		{Garc\'{\i}a\ensuremath{-}Patr\'on}}]{Quesada_exact_simulation_speedup}%
	\BibitemOpen
	\bibfield  {author} {\bibinfo {author} {\bibfnamefont {N.}~\bibnamefont
			{Quesada}}, \bibinfo {author} {\bibfnamefont {R.~S.}\ \bibnamefont
			{Chadwick}}, \bibinfo {author} {\bibfnamefont {B.~A.}\ \bibnamefont {Bell}},
		\bibinfo {author} {\bibfnamefont {J.~M.}\ \bibnamefont {Arrazola}}, \bibinfo
		{author} {\bibfnamefont {T.}~\bibnamefont {Vincent}}, \bibinfo {author}
		{\bibfnamefont {H.}~\bibnamefont {Qi}},\ and\ \bibinfo {author}
		{\bibfnamefont {R.}~\bibnamefont {Garc\'{\i}a\ensuremath{-}Patr\'on}},\
	}\bibfield  {title} {\bibinfo {title} {Quadratic speed-up for simulating
			gaussian boson sampling},\ }\href
	{https://doi.org/10.1103/PRXQuantum.3.010306} {\bibfield  {journal} {\bibinfo
			{journal} {PRX Quantum}\ }\textbf {\bibinfo {volume} {3}},\ \bibinfo {pages}
		{010306} (\bibinfo {year} {2022})}\BibitemShut {NoStop}%
	\bibitem [{\citenamefont {Bulmer}\ \emph {et~al.}(2022)\citenamefont {Bulmer},
		\citenamefont {Bell}, \citenamefont {Chadwick}, \citenamefont {Jones},
		\citenamefont {Moise}, \citenamefont {Rigazzi}, \citenamefont {Thorbecke},
		\citenamefont {Haus}, \citenamefont {Vaerenbergh}, \citenamefont {Patel},
		\citenamefont {Walmsley},\ and\ \citenamefont {Laing}}]{Bulmer_Markov_GBS}%
	\BibitemOpen
	\bibfield  {author} {\bibinfo {author} {\bibfnamefont {J.~F.~F.}\
			\bibnamefont {Bulmer}}, \bibinfo {author} {\bibfnamefont {B.~A.}\
			\bibnamefont {Bell}}, \bibinfo {author} {\bibfnamefont {R.~S.}\ \bibnamefont
			{Chadwick}}, \bibinfo {author} {\bibfnamefont {A.~E.}\ \bibnamefont {Jones}},
		\bibinfo {author} {\bibfnamefont {D.}~\bibnamefont {Moise}}, \bibinfo
		{author} {\bibfnamefont {A.}~\bibnamefont {Rigazzi}}, \bibinfo {author}
		{\bibfnamefont {J.}~\bibnamefont {Thorbecke}}, \bibinfo {author}
		{\bibfnamefont {U.-U.}\ \bibnamefont {Haus}}, \bibinfo {author}
		{\bibfnamefont {T.~V.}\ \bibnamefont {Vaerenbergh}}, \bibinfo {author}
		{\bibfnamefont {R.~B.}\ \bibnamefont {Patel}}, \bibinfo {author}
		{\bibfnamefont {I.~A.}\ \bibnamefont {Walmsley}},\ and\ \bibinfo {author}
		{\bibfnamefont {A.}~\bibnamefont {Laing}},\ }\bibfield  {title} {\bibinfo
		{title} {The boundary for quantum advantage in gaussian boson sampling},\
	}\href {https://doi.org/10.1126/sciadv.abl9236} {\bibfield  {journal}
		{\bibinfo  {journal} {Science Advances}\ }\textbf {\bibinfo {volume} {8}},\
		\bibinfo {pages} {eabl9236} (\bibinfo {year} {2022})}\BibitemShut {NoStop}%
	\bibitem [{\citenamefont {Aaronson}\ and\ \citenamefont
		{Arkhipov}(2014)}]{Aaronson14}%
	\BibitemOpen
	\bibfield  {author} {\bibinfo {author} {\bibfnamefont {S.}~\bibnamefont
			{Aaronson}}\ and\ \bibinfo {author} {\bibfnamefont {A.}~\bibnamefont
			{Arkhipov}},\ }\bibfield  {title} {\bibinfo {title} {Bosonsampling is far
			from uniform},\ }\href@noop {} {\bibfield  {journal} {\bibinfo  {journal}
			{Quantum Information \& Computation}\ }\textbf {\bibinfo {volume} {14}},\
		\bibinfo {pages} {1383} (\bibinfo {year} {2014})}\BibitemShut {NoStop}%
	\bibitem [{\citenamefont {Tichy}(2015)}]{Tichy}%
	\BibitemOpen
	\bibfield  {author} {\bibinfo {author} {\bibfnamefont {M.~C.}\ \bibnamefont
			{Tichy}},\ }\bibfield  {title} {\bibinfo {title} {Sampling of partially
			distinguishable bosons and the relation to the multidimensional permanent},\
	}\href {https://doi.org/10.1103/PhysRevA.91.022316} {\bibfield  {journal}
		{\bibinfo  {journal} {Phys. Rev. A}\ }\textbf {\bibinfo {volume} {91}},\
		\bibinfo {pages} {022316} (\bibinfo {year} {2015})}\BibitemShut {NoStop}%
	\bibitem [{\citenamefont {Spagnolo}\ \emph {et~al.}(2014)\citenamefont
		{Spagnolo}, \citenamefont {Vitelli}, \citenamefont {Bentivegna},
		\citenamefont {Brod}, \citenamefont {Crespi}, \citenamefont {Flamini},
		\citenamefont {Giacomini}, \citenamefont {Milani}, \citenamefont {Ramponi},
		\citenamefont {Mataloni}, \citenamefont {Osellame}, \citenamefont
		{Galv{\~a}o},\ and\ \citenamefont {Sciarrino}}]{Spagnolo2}%
	\BibitemOpen
	\bibfield  {author} {\bibinfo {author} {\bibfnamefont {N.}~\bibnamefont
			{Spagnolo}}, \bibinfo {author} {\bibfnamefont {C.}~\bibnamefont {Vitelli}},
		\bibinfo {author} {\bibfnamefont {M.}~\bibnamefont {Bentivegna}}, \bibinfo
		{author} {\bibfnamefont {D.~J.}\ \bibnamefont {Brod}}, \bibinfo {author}
		{\bibfnamefont {A.}~\bibnamefont {Crespi}}, \bibinfo {author} {\bibfnamefont
			{F.}~\bibnamefont {Flamini}}, \bibinfo {author} {\bibfnamefont
			{S.}~\bibnamefont {Giacomini}}, \bibinfo {author} {\bibfnamefont
			{G.}~\bibnamefont {Milani}}, \bibinfo {author} {\bibfnamefont
			{R.}~\bibnamefont {Ramponi}}, \bibinfo {author} {\bibfnamefont
			{P.}~\bibnamefont {Mataloni}}, \bibinfo {author} {\bibfnamefont
			{R.}~\bibnamefont {Osellame}}, \bibinfo {author} {\bibfnamefont {E.~F.}\
			\bibnamefont {Galv{\~a}o}},\ and\ \bibinfo {author} {\bibfnamefont
			{F.}~\bibnamefont {Sciarrino}},\ }\bibfield  {title} {\bibinfo {title}
		{Experimental validation of photonic boson sampling},\ }\href
	{https://doi.org/10.1038/nphoton.2014.135} {\bibfield  {journal} {\bibinfo
			{journal} {Nature Photonics}\ }\textbf {\bibinfo {volume} {8}},\ \bibinfo
		{pages} {615} (\bibinfo {year} {2014})}\BibitemShut {NoStop}%
	\bibitem [{\citenamefont {Carolan}\ \emph {et~al.}(2015)\citenamefont
		{Carolan}, \citenamefont {Harrold}, \citenamefont {Sparrow}, \citenamefont
		{Mart{\'\i}n-L{\'o}pez}, \citenamefont {Russell}, \citenamefont
		{Silverstone}, \citenamefont {Shadbolt}, \citenamefont {Matsuda},
		\citenamefont {Oguma}, \citenamefont {Itoh}, \citenamefont {Marshall},
		\citenamefont {Thompson}, \citenamefont {Matthews}, \citenamefont
		{Hashimoto}, \citenamefont {O{\textquoteright}Brien},\ and\ \citenamefont
		{Laing}}]{Carolan15}%
	\BibitemOpen
	\bibfield  {author} {\bibinfo {author} {\bibfnamefont {J.}~\bibnamefont
			{Carolan}}, \bibinfo {author} {\bibfnamefont {C.}~\bibnamefont {Harrold}},
		\bibinfo {author} {\bibfnamefont {C.}~\bibnamefont {Sparrow}}, \bibinfo
		{author} {\bibfnamefont {E.}~\bibnamefont {Mart{\'\i}n-L{\'o}pez}}, \bibinfo
		{author} {\bibfnamefont {N.~J.}\ \bibnamefont {Russell}}, \bibinfo {author}
		{\bibfnamefont {J.~W.}\ \bibnamefont {Silverstone}}, \bibinfo {author}
		{\bibfnamefont {P.~J.}\ \bibnamefont {Shadbolt}}, \bibinfo {author}
		{\bibfnamefont {N.}~\bibnamefont {Matsuda}}, \bibinfo {author} {\bibfnamefont
			{M.}~\bibnamefont {Oguma}}, \bibinfo {author} {\bibfnamefont
			{M.}~\bibnamefont {Itoh}}, \bibinfo {author} {\bibfnamefont {G.~D.}\
			\bibnamefont {Marshall}}, \bibinfo {author} {\bibfnamefont {M.~G.}\
			\bibnamefont {Thompson}}, \bibinfo {author} {\bibfnamefont {J.~C.~F.}\
			\bibnamefont {Matthews}}, \bibinfo {author} {\bibfnamefont {T.}~\bibnamefont
			{Hashimoto}}, \bibinfo {author} {\bibfnamefont {J.~L.}\ \bibnamefont
			{O{\textquoteright}Brien}},\ and\ \bibinfo {author} {\bibfnamefont
			{A.}~\bibnamefont {Laing}},\ }\bibfield  {title} {\bibinfo {title} {Universal
			linear optics},\ }\href {https://doi.org/10.1126/science.aab3642} {\bibfield
		{journal} {\bibinfo  {journal} {Science}\ }\textbf {\bibinfo {volume}
			{349}},\ \bibinfo {pages} {711} (\bibinfo {year} {2015})}\BibitemShut
	{NoStop}%
	\bibitem [{\citenamefont {Crespi}\ \emph {et~al.}(2016)\citenamefont {Crespi},
		\citenamefont {Osellame}, \citenamefont {Ramponi}, \citenamefont
		{Bentivegna}, \citenamefont {Flamini}, \citenamefont {Spagnolo},
		\citenamefont {Viggianiello}, \citenamefont {Innocenti}, \citenamefont
		{Mataloni},\ and\ \citenamefont {Sciarrino}}]{Crespi16}%
	\BibitemOpen
	\bibfield  {author} {\bibinfo {author} {\bibfnamefont {A.}~\bibnamefont
			{Crespi}}, \bibinfo {author} {\bibfnamefont {R.}~\bibnamefont {Osellame}},
		\bibinfo {author} {\bibfnamefont {R.}~\bibnamefont {Ramponi}}, \bibinfo
		{author} {\bibfnamefont {M.}~\bibnamefont {Bentivegna}}, \bibinfo {author}
		{\bibfnamefont {F.}~\bibnamefont {Flamini}}, \bibinfo {author} {\bibfnamefont
			{N.}~\bibnamefont {Spagnolo}}, \bibinfo {author} {\bibfnamefont
			{N.}~\bibnamefont {Viggianiello}}, \bibinfo {author} {\bibfnamefont
			{L.}~\bibnamefont {Innocenti}}, \bibinfo {author} {\bibfnamefont
			{P.}~\bibnamefont {Mataloni}},\ and\ \bibinfo {author} {\bibfnamefont
			{F.}~\bibnamefont {Sciarrino}},\ }\bibfield  {title} {\bibinfo {title}
		{Suppression law of quantum states in a 3d photonic fast fourier transform
			chip},\ }\href {https://doi.org/10.1038/ncomms10469} {\bibfield  {journal}
		{\bibinfo  {journal} {Nature Communications}\ }\textbf {\bibinfo {volume}
			{7}},\ \bibinfo {pages} {10469} (\bibinfo {year} {2016})}\BibitemShut
	{NoStop}%
	\bibitem [{\citenamefont {Viggianiello}\ \emph
		{et~al.}(2018{\natexlab{a}})\citenamefont {Viggianiello}, \citenamefont
		{Flamini}, \citenamefont {Innocenti}, \citenamefont {Cozzolino},
		\citenamefont {Bentivegna}, \citenamefont {Spagnolo}, \citenamefont {Crespi},
		\citenamefont {Brod}, \citenamefont {Galv{\~{a}}o}, \citenamefont
		{Osellame},\ and\ \citenamefont {Sciarrino}}]{Viggianiello18}%
	\BibitemOpen
	\bibfield  {author} {\bibinfo {author} {\bibfnamefont {N.}~\bibnamefont
			{Viggianiello}}, \bibinfo {author} {\bibfnamefont {F.}~\bibnamefont
			{Flamini}}, \bibinfo {author} {\bibfnamefont {L.}~\bibnamefont {Innocenti}},
		\bibinfo {author} {\bibfnamefont {D.}~\bibnamefont {Cozzolino}}, \bibinfo
		{author} {\bibfnamefont {M.}~\bibnamefont {Bentivegna}}, \bibinfo {author}
		{\bibfnamefont {N.}~\bibnamefont {Spagnolo}}, \bibinfo {author}
		{\bibfnamefont {A.}~\bibnamefont {Crespi}}, \bibinfo {author} {\bibfnamefont
			{D.~J.}\ \bibnamefont {Brod}}, \bibinfo {author} {\bibfnamefont {E.~F.}\
			\bibnamefont {Galv{\~{a}}o}}, \bibinfo {author} {\bibfnamefont
			{R.}~\bibnamefont {Osellame}},\ and\ \bibinfo {author} {\bibfnamefont
			{F.}~\bibnamefont {Sciarrino}},\ }\bibfield  {title} {\bibinfo {title}
		{Experimental generalized quantum suppression law in sylvester
			interferometers},\ }\href {https://doi.org/10.1088/1367-2630/aaad92}
	{\bibfield  {journal} {\bibinfo  {journal} {New J. Phys.}\ }\textbf {\bibinfo
			{volume} {20}},\ \bibinfo {pages} {033017} (\bibinfo {year}
		{2018}{\natexlab{a}})}\BibitemShut {NoStop}%
	\bibitem [{\citenamefont {Walschaers}\ \emph {et~al.}(2016)\citenamefont
		{Walschaers}, \citenamefont {Kuipers}, \citenamefont {Urbina}, \citenamefont
		{Mayer}, \citenamefont {Tichy}, \citenamefont {Richter},\ and\ \citenamefont
		{Buchleitner}}]{Walschaers16}%
	\BibitemOpen
	\bibfield  {author} {\bibinfo {author} {\bibfnamefont {M.}~\bibnamefont
			{Walschaers}}, \bibinfo {author} {\bibfnamefont {J.}~\bibnamefont {Kuipers}},
		\bibinfo {author} {\bibfnamefont {J.-D.}\ \bibnamefont {Urbina}}, \bibinfo
		{author} {\bibfnamefont {K.}~\bibnamefont {Mayer}}, \bibinfo {author}
		{\bibfnamefont {M.~C.}\ \bibnamefont {Tichy}}, \bibinfo {author}
		{\bibfnamefont {K.}~\bibnamefont {Richter}},\ and\ \bibinfo {author}
		{\bibfnamefont {A.}~\bibnamefont {Buchleitner}},\ }\bibfield  {title}
	{\bibinfo {title} {Statistical benchmark for {BosonSampling}},\ }\href
	{https://doi.org/10.1088/1367-2630/18/3/032001} {\bibfield  {journal}
		{\bibinfo  {journal} {New Journal of Physics}\ }\textbf {\bibinfo {volume}
			{18}},\ \bibinfo {pages} {032001} (\bibinfo {year} {2016})}\BibitemShut
	{NoStop}%
	\bibitem [{\citenamefont {Giordani}\ \emph {et~al.}(2018)\citenamefont
		{Giordani}, \citenamefont {Flamini}, \citenamefont {Pompili}, \citenamefont
		{Viggianiello}, \citenamefont {Spagnolo}, \citenamefont {Crespi},
		\citenamefont {Osellame}, \citenamefont {Wiebe}, \citenamefont {Walschaers},
		\citenamefont {Buchleitner},\ and\ \citenamefont {Sciarrino}}]{Giordani18}%
	\BibitemOpen
	\bibfield  {author} {\bibinfo {author} {\bibfnamefont {T.}~\bibnamefont
			{Giordani}}, \bibinfo {author} {\bibfnamefont {F.}~\bibnamefont {Flamini}},
		\bibinfo {author} {\bibfnamefont {M.}~\bibnamefont {Pompili}}, \bibinfo
		{author} {\bibfnamefont {N.}~\bibnamefont {Viggianiello}}, \bibinfo {author}
		{\bibfnamefont {N.}~\bibnamefont {Spagnolo}}, \bibinfo {author}
		{\bibfnamefont {A.}~\bibnamefont {Crespi}}, \bibinfo {author} {\bibfnamefont
			{R.}~\bibnamefont {Osellame}}, \bibinfo {author} {\bibfnamefont
			{N.}~\bibnamefont {Wiebe}}, \bibinfo {author} {\bibfnamefont
			{M.}~\bibnamefont {Walschaers}}, \bibinfo {author} {\bibfnamefont
			{A.}~\bibnamefont {Buchleitner}},\ and\ \bibinfo {author} {\bibfnamefont
			{F.}~\bibnamefont {Sciarrino}},\ }\bibfield  {title} {\bibinfo {title}
		{Experimental statistical signature of many-body quantum interference},\
	}\href {https://doi.org/10.1038/s41566-018-0097-4} {\bibfield  {journal}
		{\bibinfo  {journal} {Nature Photonics}\ }\textbf {\bibinfo {volume} {12}},\
		\bibinfo {pages} {173} (\bibinfo {year} {2018})}\BibitemShut {NoStop}%
	\bibitem [{\citenamefont {Agresti}\ \emph {et~al.}(2019)\citenamefont
		{Agresti}, \citenamefont {Viggianiello}, \citenamefont {Flamini},
		\citenamefont {Spagnolo}, \citenamefont {Crespi}, \citenamefont {Osellame},
		\citenamefont {Wiebe},\ and\ \citenamefont {Sciarrino}}]{agresti2019pattern}%
	\BibitemOpen
	\bibfield  {author} {\bibinfo {author} {\bibfnamefont {I.}~\bibnamefont
			{Agresti}}, \bibinfo {author} {\bibfnamefont {N.}~\bibnamefont
			{Viggianiello}}, \bibinfo {author} {\bibfnamefont {F.}~\bibnamefont
			{Flamini}}, \bibinfo {author} {\bibfnamefont {N.}~\bibnamefont {Spagnolo}},
		\bibinfo {author} {\bibfnamefont {A.}~\bibnamefont {Crespi}}, \bibinfo
		{author} {\bibfnamefont {R.}~\bibnamefont {Osellame}}, \bibinfo {author}
		{\bibfnamefont {N.}~\bibnamefont {Wiebe}},\ and\ \bibinfo {author}
		{\bibfnamefont {F.}~\bibnamefont {Sciarrino}},\ }\bibfield  {title} {\bibinfo
		{title} {Pattern recognition techniques for boson sampling validation},\
	}\href {https://doi.org/10.1103/PhysRevX.9.011013} {\bibfield  {journal}
		{\bibinfo  {journal} {Phys. Rev. X}\ }\textbf {\bibinfo {volume} {9}},\
		\bibinfo {pages} {011013} (\bibinfo {year} {2019})}\BibitemShut {NoStop}%
	\bibitem [{\citenamefont {Flamini}\ \emph {et~al.}(2019)\citenamefont
		{Flamini}, \citenamefont {Spagnolo},\ and\ \citenamefont
		{Sciarrino}}]{FlaminiTSNE}%
	\BibitemOpen
	\bibfield  {author} {\bibinfo {author} {\bibfnamefont {F.}~\bibnamefont
			{Flamini}}, \bibinfo {author} {\bibfnamefont {N.}~\bibnamefont {Spagnolo}},\
		and\ \bibinfo {author} {\bibfnamefont {F.}~\bibnamefont {Sciarrino}},\
	}\bibfield  {title} {\bibinfo {title} {Visual assessment of multi-photon
			interference},\ }\href {https://doi.org/10.1088/2058-9565/ab04fc} {\bibfield
		{journal} {\bibinfo  {journal} {Quantum Science and Technology}\ }\textbf
		{\bibinfo {volume} {4}},\ \bibinfo {pages} {024008} (\bibinfo {year}
		{2019})}\BibitemShut {NoStop}%
	\bibitem [{\citenamefont {Viggianiello}\ \emph
		{et~al.}(2018{\natexlab{b}})\citenamefont {Viggianiello}, \citenamefont
		{Flamini}, \citenamefont {Bentivegna}, \citenamefont {Spagnolo},
		\citenamefont {Crespi}, \citenamefont {Brod}, \citenamefont {Galvão},
		\citenamefont {Osellame},\ and\ \citenamefont
		{Sciarrino}}]{Viggianiello17optimal}%
	\BibitemOpen
	\bibfield  {author} {\bibinfo {author} {\bibfnamefont {N.}~\bibnamefont
			{Viggianiello}}, \bibinfo {author} {\bibfnamefont {F.}~\bibnamefont
			{Flamini}}, \bibinfo {author} {\bibfnamefont {M.}~\bibnamefont {Bentivegna}},
		\bibinfo {author} {\bibfnamefont {N.}~\bibnamefont {Spagnolo}}, \bibinfo
		{author} {\bibfnamefont {A.}~\bibnamefont {Crespi}}, \bibinfo {author}
		{\bibfnamefont {D.~J.}\ \bibnamefont {Brod}}, \bibinfo {author}
		{\bibfnamefont {E.~F.}\ \bibnamefont {Galvão}}, \bibinfo {author}
		{\bibfnamefont {R.}~\bibnamefont {Osellame}},\ and\ \bibinfo {author}
		{\bibfnamefont {F.}~\bibnamefont {Sciarrino}},\ }\bibfield  {title} {\bibinfo
		{title} {Optimal photonic indistinguishability tests in multimode networks},\
	}\href {https://doi.org/https://doi.org/10.1016/j.scib.2018.10.009}
	{\bibfield  {journal} {\bibinfo  {journal} {Sci. Bull.}\ }\textbf {\bibinfo
			{volume} {63}},\ \bibinfo {pages} {1470 } (\bibinfo {year}
		{2018}{\natexlab{b}})}\BibitemShut {NoStop}%
	\bibitem [{\citenamefont {Renema}(2020{\natexlab{a}})}]{Renema_partial_2020}%
	\BibitemOpen
	\bibfield  {author} {\bibinfo {author} {\bibfnamefont {J.~J.}\ \bibnamefont
			{Renema}},\ }\bibfield  {title} {\bibinfo {title} {Simulability of partially
			distinguishable superposition and gaussian boson sampling},\ }\href
	{https://doi.org/10.1103/PhysRevA.101.063840} {\bibfield  {journal} {\bibinfo
			{journal} {Phys. Rev. A}\ }\textbf {\bibinfo {volume} {101}},\ \bibinfo
		{pages} {063840} (\bibinfo {year} {2020}{\natexlab{a}})}\BibitemShut
	{NoStop}%
	\bibitem [{\citenamefont {Renema}\ \emph {et~al.}(2018)\citenamefont {Renema},
		\citenamefont {Menssen}, \citenamefont {Clements}, \citenamefont {Triginer},
		\citenamefont {Kolthammer},\ and\ \citenamefont
		{Walmsley}}]{Renema_2018_classical}%
	\BibitemOpen
	\bibfield  {author} {\bibinfo {author} {\bibfnamefont {J.~J.}\ \bibnamefont
			{Renema}}, \bibinfo {author} {\bibfnamefont {A.}~\bibnamefont {Menssen}},
		\bibinfo {author} {\bibfnamefont {W.~R.}\ \bibnamefont {Clements}}, \bibinfo
		{author} {\bibfnamefont {G.}~\bibnamefont {Triginer}}, \bibinfo {author}
		{\bibfnamefont {W.~S.}\ \bibnamefont {Kolthammer}},\ and\ \bibinfo {author}
		{\bibfnamefont {I.~A.}\ \bibnamefont {Walmsley}},\ }\bibfield  {title}
	{\bibinfo {title} {Efficient classical algorithm for boson sampling with
			partially distinguishable photons},\ }\href
	{https://doi.org/10.1103/PhysRevLett.120.220502} {\bibfield  {journal}
		{\bibinfo  {journal} {Phys. Rev. Lett.}\ }\textbf {\bibinfo {volume} {120}},\
		\bibinfo {pages} {220502} (\bibinfo {year} {2018})}\BibitemShut {NoStop}%
	\bibitem [{\citenamefont {Moylett}\ \emph {et~al.}(2019)\citenamefont
		{Moylett}, \citenamefont {Garc{\'{\i}}a-Patr{\'{o}}n}, \citenamefont
		{Renema},\ and\ \citenamefont {Turner}}]{Moylett_2019}%
	\BibitemOpen
	\bibfield  {author} {\bibinfo {author} {\bibfnamefont {A.~E.}\ \bibnamefont
			{Moylett}}, \bibinfo {author} {\bibfnamefont {R.}~\bibnamefont
			{Garc{\'{\i}}a-Patr{\'{o}}n}}, \bibinfo {author} {\bibfnamefont {J.~J.}\
			\bibnamefont {Renema}},\ and\ \bibinfo {author} {\bibfnamefont {P.~S.}\
			\bibnamefont {Turner}},\ }\bibfield  {title} {\bibinfo {title} {Classically
			simulating near-term partially-distinguishable and lossy boson sampling},\
	}\href {https://doi.org/10.1088/2058-9565/ab5555} {\bibfield  {journal}
		{\bibinfo  {journal} {Quantum Science and Technology}\ }\textbf {\bibinfo
			{volume} {5}},\ \bibinfo {pages} {015001} (\bibinfo {year}
		{2019})}\BibitemShut {NoStop}%
	\bibitem [{\citenamefont {Phillips}\ \emph {et~al.}(2019)\citenamefont
		{Phillips}, \citenamefont {Walschaers}, \citenamefont {Renema}, \citenamefont
		{Walmsley}, \citenamefont {Treps},\ and\ \citenamefont {Sperling}}]{hanbury}%
	\BibitemOpen
	\bibfield  {author} {\bibinfo {author} {\bibfnamefont {D.~S.}\ \bibnamefont
			{Phillips}}, \bibinfo {author} {\bibfnamefont {M.}~\bibnamefont
			{Walschaers}}, \bibinfo {author} {\bibfnamefont {J.~J.}\ \bibnamefont
			{Renema}}, \bibinfo {author} {\bibfnamefont {I.~A.}\ \bibnamefont
			{Walmsley}}, \bibinfo {author} {\bibfnamefont {N.}~\bibnamefont {Treps}},\
		and\ \bibinfo {author} {\bibfnamefont {J.}~\bibnamefont {Sperling}},\
	}\bibfield  {title} {\bibinfo {title} {Benchmarking of gaussian boson
			sampling using two-point correlators},\ }\href
	{https://doi.org/10.1103/PhysRevA.99.023836} {\bibfield  {journal} {\bibinfo
			{journal} {Phys. Rev. A}\ }\textbf {\bibinfo {volume} {99}},\ \bibinfo
		{pages} {023836} (\bibinfo {year} {2019})}\BibitemShut {NoStop}%
	\bibitem [{\citenamefont {Popova}\ and\ \citenamefont
		{Rubtsov}(2021)}]{popova2021cracking}%
	\BibitemOpen
	\bibfield  {author} {\bibinfo {author} {\bibfnamefont {A.~S.}\ \bibnamefont
			{Popova}}\ and\ \bibinfo {author} {\bibfnamefont {A.~N.}\ \bibnamefont
			{Rubtsov}},\ }\href@noop {} {\bibinfo {title} {Cracking the quantum advantage
			threshold for gaussian boson sampling}} (\bibinfo {year} {2021}),\ \Eprint
	{https://arxiv.org/abs/2106.01445} {arXiv:2106.01445 [quant-ph]} \BibitemShut
	{NoStop}%
	\bibitem [{\citenamefont {Renema}(2020{\natexlab{b}})}]{renema2020marginal}%
	\BibitemOpen
	\bibfield  {author} {\bibinfo {author} {\bibfnamefont {J.~J.}\ \bibnamefont
			{Renema}},\ }\href@noop {} {\bibinfo {title} {Marginal probabilities in boson
			samplers with arbitrary input states}} (\bibinfo {year}
	{2020}{\natexlab{b}}),\ \Eprint {https://arxiv.org/abs/2012.14917}
	{arXiv:2012.14917 [quant-ph]} \BibitemShut {NoStop}%
	\bibitem [{\citenamefont {Villalonga}\ \emph {et~al.}(2021)\citenamefont
		{Villalonga}, \citenamefont {Niu}, \citenamefont {Li}, \citenamefont {Neven},
		\citenamefont {Platt}, \citenamefont {Smelyanskiy},\ and\ \citenamefont
		{Boixo}}]{villalonga2021efficient}%
	\BibitemOpen
	\bibfield  {author} {\bibinfo {author} {\bibfnamefont {B.}~\bibnamefont
			{Villalonga}}, \bibinfo {author} {\bibfnamefont {M.~Y.}\ \bibnamefont {Niu}},
		\bibinfo {author} {\bibfnamefont {L.}~\bibnamefont {Li}}, \bibinfo {author}
		{\bibfnamefont {H.}~\bibnamefont {Neven}}, \bibinfo {author} {\bibfnamefont
			{J.~C.}\ \bibnamefont {Platt}}, \bibinfo {author} {\bibfnamefont {V.~N.}\
			\bibnamefont {Smelyanskiy}},\ and\ \bibinfo {author} {\bibfnamefont
			{S.}~\bibnamefont {Boixo}},\ }\href@noop {} {\bibinfo {title} {Efficient
			approximation of experimental gaussian boson sampling}} (\bibinfo {year}
	{2021}),\ \Eprint {https://arxiv.org/abs/2109.11525} {arXiv:2109.11525
		[quant-ph]} \BibitemShut {NoStop}%
	\bibitem [{\citenamefont {Oszmaniec}\ and\ \citenamefont
		{Brod}(2018)}]{Oszmaniec_2018}%
	\BibitemOpen
	\bibfield  {author} {\bibinfo {author} {\bibfnamefont {M.}~\bibnamefont
			{Oszmaniec}}\ and\ \bibinfo {author} {\bibfnamefont {D.~J.}\ \bibnamefont
			{Brod}},\ }\bibfield  {title} {\bibinfo {title} {Classical simulation of
			photonic linear optics with lost particles},\ }\href
	{https://doi.org/10.1088/1367-2630/aadfa8} {\bibfield  {journal} {\bibinfo
			{journal} {New Journal of Physics}\ }\textbf {\bibinfo {volume} {20}},\
		\bibinfo {pages} {092002} (\bibinfo {year} {2018})}\BibitemShut {NoStop}%
	\bibitem [{\citenamefont {Garc{\'{i}}a-Patr{\'{o}}n}\ \emph
		{et~al.}(2019)\citenamefont {Garc{\'{i}}a-Patr{\'{o}}n}, \citenamefont
		{Renema},\ and\ \citenamefont
		{Shchesnovich}}]{GarciaPatron2019simulatingboson}%
	\BibitemOpen
	\bibfield  {author} {\bibinfo {author} {\bibfnamefont {R.}~\bibnamefont
			{Garc{\'{i}}a-Patr{\'{o}}n}}, \bibinfo {author} {\bibfnamefont {J.~J.}\
			\bibnamefont {Renema}},\ and\ \bibinfo {author} {\bibfnamefont
			{V.}~\bibnamefont {Shchesnovich}},\ }\bibfield  {title} {\bibinfo {title}
		{Simulating boson sampling in lossy architectures},\ }\href
	{https://doi.org/10.22331/q-2019-08-05-169} {\bibfield  {journal} {\bibinfo
			{journal} {{Quantum}}\ }\textbf {\bibinfo {volume} {3}},\ \bibinfo {pages}
		{169} (\bibinfo {year} {2019})}\BibitemShut {NoStop}%
	\bibitem [{\citenamefont {Brod}\ and\ \citenamefont
		{Oszmaniec}(2020)}]{Brod2020classicalsimulation}%
	\BibitemOpen
	\bibfield  {author} {\bibinfo {author} {\bibfnamefont {D.~J.}\ \bibnamefont
			{Brod}}\ and\ \bibinfo {author} {\bibfnamefont {M.}~\bibnamefont
			{Oszmaniec}},\ }\bibfield  {title} {\bibinfo {title} {Classical simulation of
			linear optics subject to nonuniform losses},\ }\href
	{https://doi.org/10.22331/q-2020-05-14-267} {\bibfield  {journal} {\bibinfo
			{journal} {{Quantum}}\ }\textbf {\bibinfo {volume} {4}},\ \bibinfo {pages}
		{267} (\bibinfo {year} {2020})}\BibitemShut {NoStop}%
	\bibitem [{\citenamefont {Qi}\ \emph {et~al.}(2020)\citenamefont {Qi},
		\citenamefont {Brod}, \citenamefont {Quesada},\ and\ \citenamefont
		{Garc\'{\i}a-Patr\'on}}]{Qi_lossyGBS}%
	\BibitemOpen
	\bibfield  {author} {\bibinfo {author} {\bibfnamefont {H.}~\bibnamefont
			{Qi}}, \bibinfo {author} {\bibfnamefont {D.~J.}\ \bibnamefont {Brod}},
		\bibinfo {author} {\bibfnamefont {N.}~\bibnamefont {Quesada}},\ and\ \bibinfo
		{author} {\bibfnamefont {R.}~\bibnamefont {Garc\'{\i}a-Patr\'on}},\
	}\bibfield  {title} {\bibinfo {title} {Regimes of classical simulability for
			noisy gaussian boson sampling},\ }\href
	{https://doi.org/10.1103/PhysRevLett.124.100502} {\bibfield  {journal}
		{\bibinfo  {journal} {Phys. Rev. Lett.}\ }\textbf {\bibinfo {volume} {124}},\
		\bibinfo {pages} {100502} (\bibinfo {year} {2020})}\BibitemShut {NoStop}%
	\bibitem [{\citenamefont {Kruse}\ \emph {et~al.}(2019)\citenamefont {Kruse},
		\citenamefont {Hamilton}, \citenamefont {Sansoni}, \citenamefont {Barkhofen},
		\citenamefont {Silberhorn},\ and\ \citenamefont {Jex}}]{DetailedstudyGBS}%
	\BibitemOpen
	\bibfield  {author} {\bibinfo {author} {\bibfnamefont {R.}~\bibnamefont
			{Kruse}}, \bibinfo {author} {\bibfnamefont {C.~S.}\ \bibnamefont {Hamilton}},
		\bibinfo {author} {\bibfnamefont {L.}~\bibnamefont {Sansoni}}, \bibinfo
		{author} {\bibfnamefont {S.}~\bibnamefont {Barkhofen}}, \bibinfo {author}
		{\bibfnamefont {C.}~\bibnamefont {Silberhorn}},\ and\ \bibinfo {author}
		{\bibfnamefont {I.}~\bibnamefont {Jex}},\ }\bibfield  {title} {\bibinfo
		{title} {Detailed study of gaussian boson sampling},\ }\href
	{https://doi.org/10.1103/PhysRevA.100.032326} {\bibfield  {journal} {\bibinfo
			{journal} {Phys. Rev. A}\ }\textbf {\bibinfo {volume} {100}},\ \bibinfo
		{pages} {032326} (\bibinfo {year} {2019})}\BibitemShut {NoStop}%
	\bibitem [{\citenamefont {Caianiello}(1953)}]{Caianiello1953}%
	\BibitemOpen
	\bibfield  {author} {\bibinfo {author} {\bibfnamefont {E.~R.}\ \bibnamefont
			{Caianiello}},\ }\bibfield  {title} {\bibinfo {title} {On quantum field
			theory --- i: explicit solution of dyson's equation in electrodynamics
			without use of feynman graphs},\ }\href {https://doi.org/10.1007/BF02781659}
	{\bibfield  {journal} {\bibinfo  {journal} {Il Nuovo Cimento (1943-1954)}\
		}\textbf {\bibinfo {volume} {10}},\ \bibinfo {pages} {1634} (\bibinfo {year}
		{1953})}\BibitemShut {NoStop}%
	\bibitem [{\citenamefont {Arrazola}\ \emph {et~al.}(2018)\citenamefont
		{Arrazola}, \citenamefont {Bromley},\ and\ \citenamefont
		{Rebentrost}}]{ArrazolaQOpt}%
	\BibitemOpen
	\bibfield  {author} {\bibinfo {author} {\bibfnamefont {J.~M.}\ \bibnamefont
			{Arrazola}}, \bibinfo {author} {\bibfnamefont {T.~R.}\ \bibnamefont
			{Bromley}},\ and\ \bibinfo {author} {\bibfnamefont {P.}~\bibnamefont
			{Rebentrost}},\ }\bibfield  {title} {\bibinfo {title} {Quantum approximate
			optimization with gaussian boson sampling},\ }\href
	{https://doi.org/10.1103/PhysRevA.98.012322} {\bibfield  {journal} {\bibinfo
			{journal} {Phys. Rev. A}\ }\textbf {\bibinfo {volume} {98}},\ \bibinfo
		{pages} {012322} (\bibinfo {year} {2018})}\BibitemShut {NoStop}%
	\bibitem [{\citenamefont {Killoran}\ \emph {et~al.}(2019)\citenamefont
		{Killoran}, \citenamefont {Izaac}, \citenamefont {Quesada}, \citenamefont
		{Bergholm}, \citenamefont {Amy},\ and\ \citenamefont
		{Weedbrook}}]{Killoran2019strawberryfields}%
	\BibitemOpen
	\bibfield  {author} {\bibinfo {author} {\bibfnamefont {N.}~\bibnamefont
			{Killoran}}, \bibinfo {author} {\bibfnamefont {J.}~\bibnamefont {Izaac}},
		\bibinfo {author} {\bibfnamefont {N.}~\bibnamefont {Quesada}}, \bibinfo
		{author} {\bibfnamefont {V.}~\bibnamefont {Bergholm}}, \bibinfo {author}
		{\bibfnamefont {M.}~\bibnamefont {Amy}},\ and\ \bibinfo {author}
		{\bibfnamefont {C.}~\bibnamefont {Weedbrook}},\ }\bibfield  {title} {\bibinfo
		{title} {Strawberry {F}ields: {A} {S}oftware {P}latform for {P}hotonic
			{Q}uantum {C}omputing},\ }\href {https://doi.org/10.22331/q-2019-03-11-129}
	{\bibfield  {journal} {\bibinfo  {journal} {{Quantum}}\ }\textbf {\bibinfo
			{volume} {3}},\ \bibinfo {pages} {129} (\bibinfo {year} {2019})}\BibitemShut
	{NoStop}%
	\bibitem [{\citenamefont {Bromley}\ \emph {et~al.}(2020)\citenamefont
		{Bromley}, \citenamefont {Arrazola}, \citenamefont {Jahangiri}, \citenamefont
		{Izaac}, \citenamefont {Quesada}, \citenamefont {Gran}, \citenamefont
		{Schuld}, \citenamefont {Swinarton}, \citenamefont {Zabaneh},\ and\
		\citenamefont {Killoran}}]{Bromley_2020}%
	\BibitemOpen
	\bibfield  {author} {\bibinfo {author} {\bibfnamefont {T.~R.}\ \bibnamefont
			{Bromley}}, \bibinfo {author} {\bibfnamefont {J.~M.}\ \bibnamefont
			{Arrazola}}, \bibinfo {author} {\bibfnamefont {S.}~\bibnamefont {Jahangiri}},
		\bibinfo {author} {\bibfnamefont {J.}~\bibnamefont {Izaac}}, \bibinfo
		{author} {\bibfnamefont {N.}~\bibnamefont {Quesada}}, \bibinfo {author}
		{\bibfnamefont {A.~D.}\ \bibnamefont {Gran}}, \bibinfo {author}
		{\bibfnamefont {M.}~\bibnamefont {Schuld}}, \bibinfo {author} {\bibfnamefont
			{J.}~\bibnamefont {Swinarton}}, \bibinfo {author} {\bibfnamefont
			{Z.}~\bibnamefont {Zabaneh}},\ and\ \bibinfo {author} {\bibfnamefont
			{N.}~\bibnamefont {Killoran}},\ }\bibfield  {title} {\bibinfo {title}
		{Applications of near-term photonic quantum computers: software and
			algorithms},\ }\href {https://doi.org/10.1088/2058-9565/ab8504} {\bibfield
		{journal} {\bibinfo  {journal} {Quantum Science and Technology}\ }\textbf
		{\bibinfo {volume} {5}},\ \bibinfo {pages} {034010} (\bibinfo {year}
		{2020})}\BibitemShut {NoStop}%
	\bibitem [{\citenamefont {Gupt}\ \emph {et~al.}(2019)\citenamefont {Gupt},
		\citenamefont {Izaac},\ and\ \citenamefont {Quesada}}]{Gupt2019}%
	\BibitemOpen
	\bibfield  {author} {\bibinfo {author} {\bibfnamefont {B.}~\bibnamefont
			{Gupt}}, \bibinfo {author} {\bibfnamefont {J.}~\bibnamefont {Izaac}},\ and\
		\bibinfo {author} {\bibfnamefont {N.}~\bibnamefont {Quesada}},\ }\bibfield
	{title} {\bibinfo {title} {The walrus: a library for the calculation of
			hafnians, hermite polynomials and gaussian boson sampling},\ }\href
	{https://doi.org/10.21105/joss.01705} {\bibfield  {journal} {\bibinfo
			{journal} {Journal of Open Source Software}\ }\textbf {\bibinfo {volume}
			{4}},\ \bibinfo {pages} {1705} (\bibinfo {year} {2019})}\BibitemShut
	{NoStop}%
	\bibitem [{\citenamefont {Taballione}\ \emph {et~al.}(2021)\citenamefont
		{Taballione}, \citenamefont {van~der Meer}, \citenamefont {Snijders},
		\citenamefont {Hooijschuur}, \citenamefont {Epping}, \citenamefont
		{de~Goede}, \citenamefont {Kassenberg}, \citenamefont {Venderbosch},
		\citenamefont {Toebes}, \citenamefont {van~den Vlekkert}, \citenamefont
		{Pinkse},\ and\ \citenamefont {Renema}}]{Taballione_2021}%
	\BibitemOpen
	\bibfield  {author} {\bibinfo {author} {\bibfnamefont {C.}~\bibnamefont
			{Taballione}}, \bibinfo {author} {\bibfnamefont {R.}~\bibnamefont {van~der
				Meer}}, \bibinfo {author} {\bibfnamefont {H.~J.}\ \bibnamefont {Snijders}},
		\bibinfo {author} {\bibfnamefont {P.}~\bibnamefont {Hooijschuur}}, \bibinfo
		{author} {\bibfnamefont {J.~P.}\ \bibnamefont {Epping}}, \bibinfo {author}
		{\bibfnamefont {M.}~\bibnamefont {de~Goede}}, \bibinfo {author}
		{\bibfnamefont {B.}~\bibnamefont {Kassenberg}}, \bibinfo {author}
		{\bibfnamefont {P.}~\bibnamefont {Venderbosch}}, \bibinfo {author}
		{\bibfnamefont {C.}~\bibnamefont {Toebes}}, \bibinfo {author} {\bibfnamefont
			{H.}~\bibnamefont {van~den Vlekkert}}, \bibinfo {author} {\bibfnamefont
			{P.~W.~H.}\ \bibnamefont {Pinkse}},\ and\ \bibinfo {author} {\bibfnamefont
			{J.~J.}\ \bibnamefont {Renema}},\ }\bibfield  {title} {\bibinfo {title} {A
			universal fully reconfigurable 12-mode quantum photonic processor},\ }\href
	{https://doi.org/10.1088/2633-4356/ac168c} {\bibfield  {journal} {\bibinfo
			{journal} {Materials for Quantum Technology}\ }\textbf {\bibinfo {volume}
			{1}},\ \bibinfo {pages} {035002} (\bibinfo {year} {2021})}\BibitemShut
	{NoStop}%
	\bibitem [{\citenamefont {Hoch}\ \emph {et~al.}(2021)\citenamefont {Hoch},
		\citenamefont {Piacentini}, \citenamefont {Giordani}, \citenamefont {Tian},
		\citenamefont {Iuliano}, \citenamefont {Esposito}, \citenamefont {Camillini},
		\citenamefont {Carvacho}, \citenamefont {Ceccarelli}, \citenamefont
		{Spagnolo}, \citenamefont {Crespi}, \citenamefont {Sciarrino},\ and\
		\citenamefont {Osellame}}]{hoch2021boson}%
	\BibitemOpen
	\bibfield  {author} {\bibinfo {author} {\bibfnamefont {F.}~\bibnamefont
			{Hoch}}, \bibinfo {author} {\bibfnamefont {S.}~\bibnamefont {Piacentini}},
		\bibinfo {author} {\bibfnamefont {T.}~\bibnamefont {Giordani}}, \bibinfo
		{author} {\bibfnamefont {Z.-N.}\ \bibnamefont {Tian}}, \bibinfo {author}
		{\bibfnamefont {M.}~\bibnamefont {Iuliano}}, \bibinfo {author} {\bibfnamefont
			{C.}~\bibnamefont {Esposito}}, \bibinfo {author} {\bibfnamefont
			{A.}~\bibnamefont {Camillini}}, \bibinfo {author} {\bibfnamefont
			{G.}~\bibnamefont {Carvacho}}, \bibinfo {author} {\bibfnamefont
			{F.}~\bibnamefont {Ceccarelli}}, \bibinfo {author} {\bibfnamefont
			{N.}~\bibnamefont {Spagnolo}}, \bibinfo {author} {\bibfnamefont
			{A.}~\bibnamefont {Crespi}}, \bibinfo {author} {\bibfnamefont
			{F.}~\bibnamefont {Sciarrino}},\ and\ \bibinfo {author} {\bibfnamefont
			{R.}~\bibnamefont {Osellame}},\ }\href@noop {} {\bibinfo {title} {Boson
			sampling in a reconfigurable continuously-coupled 3d photonic circuit}}
	(\bibinfo {year} {2021}),\ \Eprint {https://arxiv.org/abs/2106.08260}
	{arXiv:2106.08260 [quant-ph]} \BibitemShut {NoStop}%
	\bibitem [{\citenamefont {Chakhmakhchyan}\ \emph {et~al.}(2017)\citenamefont
		{Chakhmakhchyan}, \citenamefont {Cerf},\ and\ \citenamefont
		{Garcia-Patron}}]{perm_pos_semi}%
	\BibitemOpen
	\bibfield  {author} {\bibinfo {author} {\bibfnamefont {L.}~\bibnamefont
			{Chakhmakhchyan}}, \bibinfo {author} {\bibfnamefont {N.~J.}\ \bibnamefont
			{Cerf}},\ and\ \bibinfo {author} {\bibfnamefont {R.}~\bibnamefont
			{Garcia-Patron}},\ }\bibfield  {title} {\bibinfo {title} {Quantum-inspired
			algorithm for estimating the permanent of positive semidefinite matrices},\
	}\href {https://doi.org/10.1103/PhysRevA.96.022329} {\bibfield  {journal}
		{\bibinfo  {journal} {Phys. Rev. A}\ }\textbf {\bibinfo {volume} {96}},\
		\bibinfo {pages} {022329} (\bibinfo {year} {2017})}\BibitemShut {NoStop}%
	\bibitem [{\citenamefont {Glauber}(2007)}]{Glauber2007}%
	\BibitemOpen
	\bibfield  {author} {\bibinfo {author} {\bibfnamefont {R.}~\bibnamefont
			{Glauber}},\ }\href@noop {} {\emph {\bibinfo {title} {Quantum Theory of
				Optical Coherence}}}\ (\bibinfo  {publisher} {Wiley-VCH, Weinheim, Germany},\
	\bibinfo {year} {2007})\BibitemShut {NoStop}%
	\bibitem [{\citenamefont {Loudon}(1983)}]{loudon1983quantum}%
	\BibitemOpen
	\bibfield  {author} {\bibinfo {author} {\bibfnamefont {R.}~\bibnamefont
			{Loudon}},\ }\href@noop {} {\emph {\bibinfo {title} {The quantum theory of
				light}}},\ Oxford science publications\ (\bibinfo  {publisher} {Clarendon
		Press},\ \bibinfo {year} {1983})\BibitemShut {NoStop}%
	\bibitem [{\citenamefont {Ioffe}\ and\ \citenamefont
		{Szegedy}(2015)}]{Ioffe2015Feb}%
	\BibitemOpen
	\bibfield  {author} {\bibinfo {author} {\bibfnamefont {S.}~\bibnamefont
			{Ioffe}}\ and\ \bibinfo {author} {\bibfnamefont {C.}~\bibnamefont
			{Szegedy}},\ }\bibfield  {title} {\bibinfo {title} {{Batch Normalization:
				Accelerating Deep Network Training by Reducing Internal Covariate Shift}},\
	}\href {https://arxiv.org/abs/1502.03167v3} {\bibfield  {journal} {\bibinfo
			{journal} {arXiv}\ } (\bibinfo {year} {2015})},\ \Eprint
	{https://arxiv.org/abs/1502.03167} {1502.03167} \BibitemShut {NoStop}%
	\bibitem [{\citenamefont {Paszke}\ \emph {et~al.}(2019)\citenamefont {Paszke},
		\citenamefont {Gross}, \citenamefont {Massa}, \citenamefont {Lerer},
		\citenamefont {Bradbury}, \citenamefont {Chanan}, \citenamefont {Killeen},
		\citenamefont {Lin}, \citenamefont {Gimelshein}, \citenamefont {Antiga},
		\citenamefont {Desmaison}, \citenamefont {Kopf}, \citenamefont {Yang},
		\citenamefont {DeVito}, \citenamefont {Raison}, \citenamefont {Tejani},
		\citenamefont {Chilamkurthy}, \citenamefont {Steiner}, \citenamefont {Fang},
		\citenamefont {Bai},\ and\ \citenamefont {Chintala}}]{pytorch}%
	\BibitemOpen
	\bibfield  {author} {\bibinfo {author} {\bibfnamefont {A.}~\bibnamefont
			{Paszke}}, \bibinfo {author} {\bibfnamefont {S.}~\bibnamefont {Gross}},
		\bibinfo {author} {\bibfnamefont {F.}~\bibnamefont {Massa}}, \bibinfo
		{author} {\bibfnamefont {A.}~\bibnamefont {Lerer}}, \bibinfo {author}
		{\bibfnamefont {J.}~\bibnamefont {Bradbury}}, \bibinfo {author}
		{\bibfnamefont {G.}~\bibnamefont {Chanan}}, \bibinfo {author} {\bibfnamefont
			{T.}~\bibnamefont {Killeen}}, \bibinfo {author} {\bibfnamefont
			{Z.}~\bibnamefont {Lin}}, \bibinfo {author} {\bibfnamefont {N.}~\bibnamefont
			{Gimelshein}}, \bibinfo {author} {\bibfnamefont {L.}~\bibnamefont {Antiga}},
		\bibinfo {author} {\bibfnamefont {A.}~\bibnamefont {Desmaison}}, \bibinfo
		{author} {\bibfnamefont {A.}~\bibnamefont {Kopf}}, \bibinfo {author}
		{\bibfnamefont {E.}~\bibnamefont {Yang}}, \bibinfo {author} {\bibfnamefont
			{Z.}~\bibnamefont {DeVito}}, \bibinfo {author} {\bibfnamefont
			{M.}~\bibnamefont {Raison}}, \bibinfo {author} {\bibfnamefont
			{A.}~\bibnamefont {Tejani}}, \bibinfo {author} {\bibfnamefont
			{S.}~\bibnamefont {Chilamkurthy}}, \bibinfo {author} {\bibfnamefont
			{B.}~\bibnamefont {Steiner}}, \bibinfo {author} {\bibfnamefont
			{L.}~\bibnamefont {Fang}}, \bibinfo {author} {\bibfnamefont {J.}~\bibnamefont
			{Bai}},\ and\ \bibinfo {author} {\bibfnamefont {S.}~\bibnamefont
			{Chintala}},\ }\bibfield  {title} {\bibinfo {title} {Pytorch: An imperative
			style, high-performance deep learning library},\ }in\ \href@noop {} {\emph
		{\bibinfo {booktitle} {Advances in Neural Information Processing Systems
				32}}}\ (\bibinfo {year} {2019})\ pp.\ \bibinfo {pages}
	{8024--8035}\BibitemShut {NoStop}%
\end{thebibliography}
\end{document}